\documentclass[aps,pre,twocolumn,eqsecnum]{revtex4}

\usepackage{amsmath,bm,epsfig,here}

  \def\Fbox#1{\vskip1ex\hbox to 8.5cm{\hfil\fboxsep0.3cm\fbox{%
    \parbox{8.0cm}{#1}}\hfil}\vskip1ex\noindent}  

\newcommand{\B}[1]{{\bm{#1}}}
\newcommand{\C}[1]{{\mathcal{#1}}}    
\newcommand{\BC}[1]{\bm{\mathcal{#1}}}
\renewcommand{\sb}[1]{_{\rm{#1}}}  
\newcommand{\Sb}[1]{_{\scriptscriptstyle\rm{#1}}} 
\newcommand{\Sp}[1]{^{\scriptscriptstyle\rm{#1}}} 

\newcommand{\eq}[1]{(\ref{#1})}
\newcommand{\Eq}[1]{Eq.~(\ref{#1})}
\newcommand{\Eqs}[1]{Eqs.~(\ref{#1})}
\newcommand{\Fig}[1]{Figure~\ref{#1}}
\newcommand{\Ref}[1]{(\ref{#1})}

\newcommand{\vK}{von-K\'arm\'an~}

\newcommand{\Rig}{\mbox{Ri}\sb{grad}}
\newcommand{\Rif}{\mbox{Ri}\sb{flux}}

\begin{document}

\title{Turbulent Fluxes in Stably Stratified   Boundary Layers }

\author{Victor S. L'vov}
    \email{victor.lvov@weizmann.ac.il} 
    \affiliation{Department of Chemical Physics, The Weizmann Institute of Science, Rehovot 76100, Israel}
    \affiliation{Department  of Theoretical Physics, Institute for Magnetism, Ukraine National Ac. of Sci., Kiev, Ukraine}

\author{Itamar Procaccia}
    \affiliation{Department of Chemical Physics, The Weizmann Institute of Science, Rehovot 76100, Israel}

\author{Oleksii Rudenko}
    \affiliation{Department of Chemical Physics, The Weizmann Institute of Science, Rehovot 76100, Israel}

\begin{abstract}
We present an extended version of an invited talk given on the International Conference ``Turbulent Mixing and Beyond". The dynamical and statistical description of stably
stratified turbulent boundary layers with the important example of
the stable atmospheric boundary layer in mind is addressed. Traditional approaches to this problem,
based on the profiles of mean quantities, velocity second-order
correlations, and dimensional estimates of the turbulent thermal
flux run into a well known difficulty,  predicting the suppression
of turbulence at a small critical value of the Richardson number, in
contradiction with observations. Phenomenological attempts to
overcome this problem suffer from various theoretical
inconsistencies. Here we present an approach taking into full
account all the second-order statistics, which allows us to respect
the conservation of total mechanical energy. The analysis culminates
in an analytic solution of the profiles of all mean quantities and
all second-order correlations  removing the unphysical predictions
of previous theories. We propose that the approach taken here is
sufficient to describe the lower parts of the atmospheric boundary
layer, as long as the Richardson number does not exceed an order of
unity. For much higher Richardson numbers the physics may change
qualitatively, requiring careful consideration of the potential
Kelvin-Helmoholtz waves and their interaction with the vortical
turbulence.
 \keywords{Atmospheric Boundary Layer, Richardson Number, Transport
  Equations, Stratification}
\end{abstract}

\maketitle


\begin{widetext}

{{
\textbf{Nomenclature}~\\ ~\\
\begin{tabular}{l c l c l c l}
  $\BC A$ & ~~ & Thermal flux production vector, (\ref{defOfA})& ~~~~ & 
  $\beta$ & ~~ & Buoyancy parameter, $\B g \widetilde \beta$ \\
  $\BC B$ & ~~ & Pressure-temperature-gradient-vector, (\ref{defs1f})& ~~~~ & 
  $\widetilde \beta$ & ~~ & Thermal expansion coefficient\\ 
  $\C C_{ij}$ & ~~ & Energy conversion tensor, (\ref{defs1B})& ~~~~ & 
  $\gamma\Sb{RI}$ & ~~ & Relaxation frequency of $\tau_{ij}$, $i = j$ \\ 
  $\C D / \C D t$ & ~~ & Substantial derivative, $\partial / \partial t + \BC U \!\cdot\! \B \nabla$ & ~~~~ & 
  $\widetilde \gamma\Sb{RI}$ & ~~ & Relaxation frequency of $\tau_{ij}$, $i \neq j$ \\ 
  $D / D t$ & ~~ & Mean substantial derivative, $\partial / \partial t + \B U \!\cdot\! \B \nabla$ & ~~~~ & 
  $\gamma\Sb{RD}$ & ~~ & Relaxation frequency of $\B F$ \\ 
  $E\Sb K$ & ~~ & Turbulent kinetic energy per unit mass, $|\B u|^2\!/2$& ~~~~ & 
  $\gamma_{uu}$ & ~~ & Relaxation frequency of $E\Sb K$ \\ 
  $E\Sb \Theta$ & ~~ & "Temperature energy" per unit mass, $\theta^2\!/2$ & ~~~~ & 
  $\gamma_{\theta\theta}$ & ~~ & Relaxation frequency of $E\Sb \Theta$ \\ 
  $\B F$ & ~~ & Turbulent thermal flux per unit mass, $ \left\langle \B u \theta \right\rangle $ & ~~~~ & 
  $\varepsilon_{ij}$ & ~~ & Dissipation tensor of $\tau_{ij}$, ($\ref{diss}$)\\ 
  $ F_*$ & ~~ & Thermal flux at zero elevation $z = 0$ & ~~~~ & 
  $\B \epsilon$ & ~~ & Dissipation vector of $\B F$, ($\ref{diss}$)\\ 
  $\B g$ & ~~ & Gravity acceleration, $\B g = -g\, \widehat{\bf{z}} $  & ~~~~ & 
  $\varepsilon$ & ~~ & Dissipation of $E\Sb \Theta$, ($\ref{diss}$)\\ 
  $L$ & ~~ & Monin-Obukhov length, $u_*^3/\beta F_*$ & ~~~~ & 
  $\Theta_d$ & ~~ & Deviation of potential temperature from BRS\\ 
  $\ell$ & ~~ & Outer scale of turbulence, external parameter & ~~~~ & 
  ${\Theta}$ & ~~ & Mean potential temperature, $ \left\langle \Theta_d \right\rangle $ \\ 
  $\C P_{ij}$ & ~~ & Rate of Reynolds stress production, ($\ref{defs1A}$) & ~~~~ & 
  $\theta$ & ~~ & Fluctuating potential temperature, $\Theta_d -\!\left\langle \Theta_d \right\rangle$  \\ 
  $p$, $\widetilde p$, $p_*$ & ~~ & Total, fluctuating and zero level pressures & ~~~~ & 
  $\theta_*$ & ~~ & Potential temperature \\ 
  Pr$\Sb T$ & ~~ & Turbulent Prandtl number, $\nu\Sb T / \chi\Sb T$ & ~~~~ & 
    & ~~ & at zero elevation, $F_* / u_*$\\ 
  Ri$\sb{flux}$ & ~~ & Flux Richardson number, $\beta F_z / \tau_{xz} S\Sb U$ & ~~~~ & 
  $\lambda_*$ & ~~ & Viscous lengthscale, $\nu / u_*$ \\ 
  Ri$\sb{grad}$ & ~~ & Gradient Richardson number, $\beta S\Sb \Theta / S^2\Sb U$ & ~~~~ & 
  $\nu$ & ~~ & Kinematic viscosity \\ 
  $\C R_{ij}$ & ~~ & Pressure-rate-of-strain-tensor, ($\ref{defs1C}$) & ~~~~ & 
  $\nu\Sb T$ & ~~ & Turbulent viscosity \\ 
  $S\Sb U$ & ~~ & Mean velocity gradient, $dU/dz$ & ~~~~ & 
  $\rho$ & ~~ & density of the fluid \\ 
  $S\Sb \Theta$ & ~~ & Mean potential temperature gradient, $d{\Theta}/dz$ & ~~~~ & 
  $\tau_{ij}$ & ~~ & Reynolds stress tensor, $ \left\langle u_i u_j \right\rangle $ \\ 
  $T$ & ~~ & Molecular temperature & ~~~~ & 
  $\tau_*$ & ~~ & Mechanical momentum flux \\ 
  $\BC U$ & ~~ & Velocity field & ~~~~ & 
    & ~~ & at zero elevation (at the ground) \\ 
  $\B U$ & ~~ & Mean velocity, $ \left\langle \BC U \right\rangle $ & ~~~~ & 
  $\chi$ & ~~ & Kinematic thermal conductivity \\ 
  $\B u$ & ~~ & Fluctuating velocity, $\BC U - \B U$ & ~~~~ & 
  $\chi\Sb T$ & ~~ & Turbulent thermal conductivity \\ 
  $u_*$ & ~~ & (Wall) friction velocity, $\sqrt{\tau_*}$ & ~~~~ & 
  BRS & ~~ & Basic Reference State \\
  $\widehat{\bf{x}} $ & ~~ & horizontal (streamwise) unit vector & ~~~~ & 
  $\widehat{\bf{z}} $ & ~~ & vertical (wall-normal)  unit vector  
\end{tabular}
}}\\
~\\
\end{widetext}

\newpage

\section*{\label{s:intro}Introduction}

The lower levels of the atmosphere are usually strongly influenced
by the Earth's surface. Known as the atmospheric boundary layer,
this is the part of the atmosphere where the surface influences the
temperature, moisture, and velocity of the air above through the
turbulent transfer of mass.

The stability of the atmospheric boundary layer depends on the
profiles of the density and the temperature as a function of the
height above the ground. During normal summer days the land mass
warms up and the temperature is higher at lower elevations. If it
were not for the decrease in density of the air as a function of the
height, such a situation of heating from below would have been
always highly unstable. In fact, the boundary layer is considered
stable as long as the temperature decreases at the dry adiabatic
lapse rate ($T'\approx - 9.8^\circ C$ per kilometer) throughout most
of the boundary layer. With such a rate of cooling one balances out
the decrease in density. With a higher degree of cooling one refers
to the atmospheric boundary layer as unstably stratified, whereas
with a lower degree of cooling the situation is stably
stratified. Stably stratified boundary layer occurs typically
during clear, calm nights. In extreme cases  turbulence tends to
cease, and radiational cooling from the surface results in a
temperature that increases with height above the surface.

The tendency of the atmosphere to be turbulent does not depend only
on the rate of cooling but also on the mean shear in the vertical
direction. The commonly used parameter to describe the tendency of
the atmosphere to be turbulent is the ``gradient" Richardson number (Richardson, 1920),
defined as
\begin{equation}\label{Rig}
\Rig \equiv  \frac{ \beta \, d\Theta (z)\big / d z}{ [d U_x/dz]^2} \,,
\end{equation}
where  $x$  is the stream-wise direction, $z$ is the height above
the ground, ${\Theta}(z)$ is the mean potential temperature
profile, (which differs from the mean temperature profile $T(z)$ by
accounting for the adiabatic cooling of the air during its
expansion: $ d{\Theta} (z)\big / d z=  d T (z)\big / d z + |T'|$), $
\beta=\widetilde \beta  g$ is the buoyancy parameter in which $\tilde \beta $ is the
adiabatic thermal expansion coefficient (for an ideal gas $\widetilde \beta= 1/ T$), and $g$ is the
gravitational acceleration. The mean shear $d U_x /d z$ is defined
in terms of the mean velocity $\B U $, which in the simplest case of
flat geometry depends only on the vertical coordinate $z$.  The
parameter $\Rig$ represents the ratio of the generation or
suppression of turbulence by buoyant production of energy to the
mechanical generation of energy by wind shear.

This paper is an extended presentation of an invited talk
 given on International Conference ``Turbulent Mixing and Beyond" devoted, in particular,  to the problems of fluid dynamics, turbulence, geophysics and  statistics, that are long-standing challenging tasks.  Here we  consider the description of stably stratified
turbulent  boundary layers (TBL), taking as an example the case of
stable thermal stratification. Since the 50's of twentieth century,
traditional models of stratified TBL generalize models of
unstratified TBL, based on the budget equations for the kinetic
energy and mechanical momentum; see reviews of Umlauf and Burchard (2005),
Weng and Taylor (2003). The main difficulty is that the budget
equations are not closed; they involve turbulent fluxes of
mechanical moments $\tau_{ij}$  (known as the ``Reynolds stress"
tensor) and a thermal flux $\B F$ (for the case of thermal
stratification): \begin{equation} \label{def3}
\tau_{ij} \equiv  \langle u_i u_j \rangle\,,\quad \B F  \equiv     \left\langle   \B u
\, \theta  \right\rangle  \,,
\end{equation} 
where $\B u$ and $\theta$ stand for the turbulent
fluctuating velocity and the potential temperature with zero mean.
The nature of the averaging procedure behind the symbol
$\langle\cdots\rangle$ will be specified below.

Earlier  estimates of the fluxes~\eq{def3} are based on the concept
of the down-gradient turbulent transport, in which, similarly to the
case of molecular transport, a flux is taken proportional to the
gradient of transported property times a corresponding (turbulent)
transport coefficient:
 \begin{subequations}\label{dt}
 \begin{eqnarray}\label{dtM}
 \tau_{xz}&=& -\nu\Sb T  {d U_x}\big /{dz}\,, \quad
\nu\Sb T \approx C_\nu \,  \ell _z \sqrt {\tau_{zz}}\,, \\
\label{dtH} 
F_z&=&- \chi \Sb T
  {d\Theta}\big /{dz}\,,\quad    ~  \chi \Sb T \approx C_\chi \, \ell _z
 \sqrt {\tau_{zz}}\,, \quad \mbox{etc.}~~~~
\end{eqnarray} \end{subequations} 
Here the turbulent-eddy viscosity $\nu \Sb T$ and turbulent thermal
conductivity $\chi \Sb T$ are estimated by dimensional reasoning via
the vertical turbulent velocity $\sqrt{\tau_{zz}}$ and a scale
$\ell_z$ (which in the simplest case is determined by the elevation
$z$). The dimensionless coefficients $C_\nu$ and $C_\chi$ are
assumed to be of the order of unity.

This approach meets serious difficulties (Zeman, 1981), in
particular, it predicts full suppression of turbulence when the
stratification exceeds a critical level, for which  $\Rig
\approx 0.25$. On the other hand, in observations of
the atmospheric turbulent boundary layer   turbulence exists for
much larger values than $\Rig=0.25$: experimentally above
$\Rig=10$ and even more (see Galperin et al. (2007) and references therein). In models for weather
predictions this problem is ``fixed" by introducing fit functions
$C_\nu (\Rig)$ and $C_\chi(\Rig)$ instead of the constant $C_\nu $
and $C_\chi$ in the model parametrization \eq{dt}. This technical
``solution" is not based on any physical derivation and just masks
the shortcomings of the model. To really solve the problem one has
to understand its physical origin, even though from a purely
formal viewpoint it is indeed possible that a dimensionless
coefficient like $C_\chi$ can be any function of $\Rig$.

To expose the physical reason for the failure of the down-gradient
approach, recall that in a stratified flow, in the presence of
gravity,  the turbulent kinetic energy is {\em not}  an integral
of motion. Only the total mechanical energy, the sum of the
kinetic and the potential energy, is conserved in the inviscid
limit. As it was shown already by Richardson, the difficult point
is that an important contribution to the potential energy comes
not just from the mean density profile, but from the density
fluctuations. Clearly, any reasonable model of  the turbulent
boundary layer must obey the conservation laws.

 The physical requirement of conserving the total mechanical
 energy calls for an explicit   consideration not only of the mean
profiles, but also of {\it all} the relevant second-order,
one-point, simultaneous correlation functions of  {\it all} the
fluctuating fields together with some closure procedure for the
appearing third order moments. First of all, in order to account for
the important effect of stratification on the anisotropy, we must
write explicit equations for the entire Reynolds stress tensor,
$\tau _{ij}= \left\langle  u_iu_j \right\rangle $ . Next, in the case of the temperature
stratified turbulent boundary layers we follow tradition [see, e.g.
 Zeeman (1981), Hunt et al. (1988), 
Schumann and Gerz (1995),  
Hanazaki and Hunt (2004), 
Keller and van Atta (2000), 
Stretch et al. (2001), 
Elperin et al. (2002), 
 Cheng et al. (2002)
Luyten et al. (2002),  and 
Rehmann and Hwang (2005)] and  account for  the turbulent potential
energy  which is proportional to the  variance of the potential
temperature deviation, $\langle \theta^2\rangle$. And last but not
least, we have to consider explicitly equations for the vertical
fluxes, $\tau_{xz}$ and $F_z$, which include the down gradient terms
proportional to the velocity and temperature gradients, and
counter-gradient terms, proportional to $F_x$ (in the equation for
$\tau_{xz}$) and to $ \left\langle  \theta^2 \right\rangle $ (in the equation for $F_z$) .

Unfortunately, the resulting second order closure seems to be
inconsistent with the variety of boundary-layer data, and many
authors took the liberty to introduce additional fitting parameters
and sometimes fitting functions to achieve a better agreement with
the data (see reviews of
 Umlauf and Burchard (2005),
 Weng and Taylor (2003),
 Zeeman, (1981),
 Melor and Yamada (1974),
  and references therein). Moreover, in the second order
closures the problem of critical Richardson number
seems to persists (Cheng et al., 2002; Canuto, 2002).

Notice that in spite of obvious inconsistency of the first-order
schemes, most of the practically used turbulent models are based
on the concept of the down-gradient transport. One of the  reasons
is that in the second-order schemes instead of two down-gradient
equations~\eq{dt} one needs to take into account eight nonlinear
coupled additional equations i.e. four equations for the Reynolds
stresses, three equations for the heat fluxes and equation for the
temperature variance. As the result, the second-order schemes have
seemed to be rather cumbersome for comprehensive analytical
treatment and have allowed to find only some relationships between
correlation functions (see, e.g., Cheng et al., 2002).
Unfortunately, the numerical solutions to the complete set of the
second-order schemes equations which involve too many fitting
parameters are much less informative in clarification of physical
picture of the phenomenon than desired analytical ones.

In this paper we suggest a relatively simple second-order closure
model of turbulent boundary layer with stable temperature
stratification that, from one hand, accounts for
main relevant physics in the stratified TBL and, from the other
hand, is simple enough to allow complete analytical treatment
including the problem of critical Ri$\sb{grad}$. To reach  this goal
we approximate the third order correlations via the first- and
second-order ones, accounting only for  the most physically
important terms. We will try to expose the approximations in a clear
and logical way, providing the physical justification as we go
along. Resulting second-order model consist of nine coupled
equations for the mean velocity and temperature gradients, four
components of the Reynolds stresses, two components of the
temperature fluxes and the temperature variance. Thanks to the
achieved simplicity of the model  we found an approximate analytical
solution of these equations, expressing all nine correlations as
functions of only one governing parameter, $\ell(z)/L$, where
$\ell(z)$ is the outer scale of turbulence (depending on the
elevation $z$ and also known as the ``dissipation scale") and $L$ --
is the Obukhov length.

 We would like also to stress, that in our approach $\ell(z)/L$
  is an external
 parameter of the problem. For small elevations
$z \ll L$, it is well accepted that $\ell(z)$ is proportional to
$z$, while the $\ell(z)$ dependence is still under debate for $z$
comparable or exceeding $L$. For $z  \gtrsim L$ the assignment and
discussion of the actual dependence of the outer scale of
turbulence, $\ell(z)$, which is manifested in the nature is out of
the scope of this paper, and is remained for future work.  At   time
being, we can analyze consequences  of our approach for  the
following versions of $\ell(z)$ dependence at $z\gg L$:  \\
\textbullet~ function $\ell(z)$ is saturated  at some level  of the
order of $L$. For concreteness we   can  take
 \begin{equation}  \label{sless}
1/\ell(z) = \sqrt{(d_1z)^{-2}+(d_2 L)^{-2}}\,,
 \quad d_1\sim d_2\sim 1\ .
 \end{equation}
\noindent \textbullet~  $\ell(z)$ is again proportional to $z$ for
elevations much larger than $L$: $\ell(z)=d_3 z$ but with the
proportionality constant $d_3< d_1$. If so,    we can also study
the case $\ell(z) \gg L$ even though such a condition may not be
realizable in Nature. In that case our analysis of the  limit
$\ell(z) \gg L$  has only a methodological character: it allows to
derive an approximate analytic solution for all the objects of
interest as functions of $\ell(z)/L$ that is also
valid for the outer scale of turbulence not exceeding a value of
the order of $L$.

It should be noticed that  traditional turbulent closures
(including ours) cannot be applied for strongly stratified flows
with $\Rig\gtrsim 1$ (may be even at  $\Rig\sim 1$). The main
reason is that these closures are roughly justified for developed
\emph{vortical} turbulence, in which the eddy-turnover time is of
the order of its life time; in other words, there are no well
defined ``quasi-particles" or waves. This is not the case for
stable stratification with $\Rig\gtrsim 1$, in which the
Brunt-V\"ais\"al\"a frequency 
\begin{equation}  \label{BVf} N\equiv \sqrt{\beta  d \Theta(z)/d\,z }\,,
 \end{equation} 
 is   larger then the eddy-turnover frequency $\gamma$.  It means that for
 $\Rig\gtrsim  1$  there are weakly decaying Kelvin-Helmoholtz internal
 gravity waves (with characteristic frequency $N$ and decay time above
$1/\gamma $), propagating on large distances, essentially effecting on
TBL, as pointed out by  Zilitinkevich, (2002). We concentrate in our paper on self-consistent
description of the lower part of the atmospheric TBL, in which
turbulence has vortical character and consequently,  large values of
$\Rig$ do not appear. We relate large values of $\Rig$ in the upper
part of TBL with contributions of the internal gravity waves  to the
energy and the energy flux in  TBL, to the momentum flux, and to the
production of (vortical) turbulent energy. Due to their instability
in a shear flow, the waves can break and create turbulent kinetic
energy. All these effects are beyond the  framework of   our
paper.  Their description in the upper  ``potential-wave" TBL and
intermediate region with the combined ``vortical-potential"
turbulent velocity field is in our nearest agenda.

To make the paper more transparent for wide audience, not
necessarily experts in atmospheric TBL, we attempt to present the
material in a self-contained manner, and organized it as follows.

In Sect.~\ref{s:balance}A we use the Oberbeck-Boussinesq
approximation and apply the standard Reynolds decomposition (into mean
values and turbulent zero-mean fluctuations of the velocity and
temperature fields) to derive equations for the mean values and
balance equations for all relevant second-order correlation
functions. In Sect. \ref{s:balance}B we demonstrate that the resulting
balance equations exactly preserve (in the non-dissipative limit)
the total mechanical energy of the system, that consists of the
kinetic energy of the mean flow, kinetic energy and potential energy
of the turbulent subsystem.

In Sect.~\ref{s:closure}   we describe the proposed closure
procedure that results in a model of stably stratified TBL, that
accounts explicitly  for all relevant second-order correlations.
The third order correlations which appear in the theory are modeled
in terms of second-order correlations in Sects.~\ref{s:closure}A and
B. Further simplifications are presented  in Sects.~\ref{s:closure}C
and D for stationary  turbulent flows in a plane geometry outside
the viscous and buffer layers.  In Sect.~\ref{s:closure}E we suggest
a generalization of the standard ``wall-normalization" to obtain
the model equations in a dimensionless form with only one governing
parameter, $\ell(z)/L$.

Section~\ref{ss:strat41} contains approximate analytical solution of
the model. It is shown that the analytical solution deviates from
the numerical counterpart in less than a few percent in the entire
interval $0\le (\ell/L)<\infty$.

The last Sect.~\ref{s:res} is devoted to a detailed description of
our results: profiles of the mean velocity and potential temperature
(Sect.~\ref{s:res}A), profiles of the turbulent kinetic and
``temperature" energies, profiles of the anisotropy of partial
kinetic energies (Sect.~\ref{s:res}B),  profiles of the turbulent
transport parameters $\nu\Sb T$ and $\chi\Sb T$, profiles of the
gradient- and flux-Richardson numbers $\Rig$ and $\Rif$, and the
dependence of the turbulent Prandtl number Pr$\Sb T$ vs. $\ell/L$ and
$\Rig$, Sect.~\ref{s:res}C. In conclusion Sect.~\ref{s:res}D, we
consider the  validity of the down-gradient transport
concept~\eq{dt} and explain why it is violated in the upper part of
TBL. The problem of critical $\Rig$ is also discussed.

\section{\label{s:balance} Simplified dynamics in a stably
 temperature-stratified TBL and their conservation laws}

The aim of this section is to consider the simplified dynamics of
a stably temperature-stratified turbulent boundary layer, aiming
finally at an explicit description of the height dependence of
important quantities like the mean velocity, mean temperature,
turbulent kinetic and potential energies, etc. In general one
expects very different profiles from those known in standard
(unstratified) wall-bounded turbulence. We want to focus on these
differences and propose that they occur already relatively close
to the ground allowing us to neglect (to the leading order) the
dependence of the density on height and the Coriolis force. We
thus begin by simplifying the hydrodynamic equations which are
used in this section.

\subsection{\label{ss:wall-units}Simplified hydrodynamic
equations and Reynolds decomposition}
First we briefly overview the derivation of the governing
equations in the Boussinesq approximation.
The system of hydrodynamic equations describing a
fluid in which the temperature is not uniform  consists of the
Navier-Stokes equations for the fluid velocity, $\BC U(\B r,t)$, a
continuity equation for the space and time dependent (total)
density of the fluid, $\rho (\B r,t)$, and of the heat balance
equation for the (total) entropy per unit mass, $\C S(\B
r,t)$,~Landau and Lifshitz, 1987.

These equations are considered with boundary conditions that
maintain the solution far from the equilibrium state, where $\BC
U=\C S=0$. These boundary conditions are $\BC U=0$ at zero
elevation, $\BC U=const$ at a high elevation of a few kilometers.
This reflects the existence of a wind at high elevation, but we do
not attempt to model the physical origin of this wind in any
detail. The only important condition with regards to this wind is
that it maintains a momentum flux towards the ground that is
prescribed as a function of the elevation. Similarly, we assume
that a stable temperature stratification is maintained such that
the heat flux towards the ground is prescribed as well.

We neglect the viscous entropy production term assuming that the
temperature gradients are large enough such that the thermal
entropy production term dominates. For simplicity of the
presentation we restrict ourselves by relatively small elevations
and disregard the Coriolis force (for more details, see Wyngaard,
1992). On the other hand we assume that the temperature and
density gradients in the entire turbulent boundary layer are
sufficiently small to allow employment of local thermodynamic
equilibrium. In other words, we assume the validity of the
equation of state.

As a ``basic reference state" (BRS) denoted hereafter by a
subscript ``$\sb b$" we use the isentropic  model of the
atmosphere, where the entropy is considered space homogeneous. Now
assuming smallness of deviations of the density and pressure
from their BRS values and exploiting the equation of state, one
obtains a simplified equation, which is already very close to the standard
Navier-Stokes equation in the Boussinesq approximation. Introducing
(generalized) potential temperature, one
results in the well-known system of hydrodynamic
equations in the Boussinesq approximation. Close to the ground, where one can neglect the dependence of the density on height, the system reads:
\begin{equation}
  \label{GovEq} 
   \frac{\C D\, \BC U }{\C D t}   =
  -\frac{\B {\nabla} p}{ \rho\sb b} - \B  \beta\, \Theta\sb {\, d}
   + \nu\, \Delta\, \BC U \,,  \  \frac{\C D\,
   \Theta\sb {\, d}}{\C D t}  = \chi\,
  \Delta\, \Theta\sb {\, d} \ .
\end{equation}
Here $ {\C D}/{\C D t}\equiv   {\partial}/{\partial{t}}+ \BC U \cdot \B
\nabla$ is the convection time derivative, $p$ -- deviation of pressure from BRS,
$\rho\sb b$ is the density in BRS, ${\bm \beta} = {\bm g} \widetilde \beta$ is the
buoyancy parameter (${\bm \beta} = -\widehat{\bm{z}} \beta$, $\beta = g \widetilde \beta$, $g$ is the gravity acceleration and $\widetilde \beta$
is the thermal expansion coefficient, which is
equal to $1/T$, reciprocal molecular temperature, for an ideal
gas), $\Theta\sb {\, d}$ is the deviation of the potential
temperature from  BRS value, $\nu$ -- kinematic viscosity and
$\chi$ is the kinematic thermal conductivity.

To develop equations
for the mean quantities and correlation functions one applies the
Reynolds decomposition: $ \BC U   =  \B U + \B u\,, \ \langle \BC
U\rangle = \B U\,, \   \left\langle \B u \right\rangle =0\,,
 \Theta\sb{\,d} = \Theta + \theta\,, \
 \langle\Theta\sb{\,d}\rangle=\Theta\,, \  \langle \theta
 \rangle= 0\,, \ p =  \langle p\rangle+\widetilde p\,, \quad \langle \widetilde p \rangle
 = 0$. Here the average $\langle \cdots \rangle$ stands for
an  averaging over a horizontal plane
at a constant elevation. This leaves the average quantities
 with a $z,t$ dependence only.
 Substituting in Eqs.~\eq{GovEq} one gets  equations of motion
 for the mean velocity and
mean temperature profiles
 \begin{equation}\label{mean}
\frac{D\, U_i}{Dt}  + \nabla\!_j\, \widetilde  \tau_{ij}=
  - \frac{ \nabla\!_i  \langle p\rangle}{\rho_b} -
\beta_i \,\Theta\,,\
\frac{D\, \Theta }{Dt}+ \B\nabla\cdot \widetilde  {\B F}=0\ .
\end{equation}
Here $ {D}/{Dt}\equiv   {\partial}/{\partial{t}}+ \B U \cdot \B \nabla$
is the mean convection derivative.  The total (molecular and
turbulent) momentum
  and thermal fluxes are
\begin{equation} \label{def-tau}\widetilde  \tau_{ij}
 \equiv   - \nu \nabla\!_j\, U_i+\tau_{ij}\,,\quad
 \widetilde  {\B F}  \equiv   -\chi \,\B\nabla  {\Theta} +\B F\,,
\end{equation}  
where $\tau_{ij} = \langle u_i u_j \rangle$ is the Reynolds stress
tensor describing the turbulent momentum flux, and $\B F = \langle
\B u \theta \rangle$ is the turbulent thermal flux.  In order to
derive equations for these correlation functions, one considers
the equations of motion for the fluctuating velocity and
temperature: 
\begin{subequations}\label{fluct}
\begin{eqnarray}   
\label{NSEFluct}
 {D\,\B u}/{D\,t}
 &\!=\!& -\B u \cdot\B  \nabla \B U -\B u \cdot\B  \nabla \B u
 +\left\langle \B u \cdot\B  \nabla \B u\right\rangle \\ \nonumber
&& -({\B \nabla \widetilde p}/{\rho_b}) + \nu\, \Delta \B u - \B
\beta\,\theta\,,\\ 
\label{fluctb}
 {D\,\theta}/{D\,t} &\!=\!& -\B
u \cdot \B \nabla  {\Theta} -\B u \cdot \B \nabla \theta +\chi\,\Delta
\theta +\left\langle \B u \cdot \B \nabla \theta \right\rangle
.~~~~~~~~
\end{eqnarray}\end{subequations} 

The whole set of the second order correlation functions includes the
Reynolds stress, $\tau_{ij}$, the turbulent thermal flux, $\B F$,
 and the ``temperature energy"
$
 E_{\theta} \equiv
 \left\langle \theta^2 \right\rangle /2$,  which is denoted and named by analogy with the
turbulent kinetic energy density (per unit mass and unit volume),
 $
 E\Sb K = \langle |\B u|^2\rangle/2=   \mbox{Tr}
 \{\tau_{ij}\}/2$.
Using \eq{fluct} one gets the following ``balance equations": 
\begin{subequations} \label{corr}
\begin{eqnarray} \label{corra}
    \frac{D\,\tau_{ij}}{D\,t} +\varepsilon_{ij}
     + \frac{\partial}{\partial{x_k}}T _{ijk}
    &=&
   \C P_{ij}- \C C_{ij} +\C \mathcal{R}_{ij}
    \,,~~~~~~~~~~~\\
\label{corrb}
  \frac{D\,F_i}{D\,t} +  \epsilon _i+
    \frac{\partial}{\partial{x_j}}T_{ij} &=& \C A_i+\C B_i
     \,,  ~~~~\\
\label{corrc}
   \frac{D\, E_{\theta}}{D\,t}+   \varepsilon+\B \nabla
   \cdot \B T&=& -\B F \cdot
    \B\nabla  {\Theta}  \ . 
\end{eqnarray}
\end{subequations}
Here we denoted the dissipations of the Reynolds-stress, heat-flux
and the temperature energy by
  \begin{eqnarray}\nonumber 
\varepsilon_{ij}&\!\!\equiv \!\!&  2\,\nu\left\langle
\frac{\partial{u_i}}{\partial{x_k}}\,
\frac{\partial{u_j}}{\partial{x_k}} \right\rangle, \
 \epsilon_{i} \equiv    \left( \nu +\chi \right) \left\langle
\frac{\partial{\theta}}{\partial{x_k}}\,
\frac{\partial{u_i}}{\partial{x_k}} \right\rangle, \\
 \label{diss}
\varepsilon & \equiv  &  \chi\, \left\langle |\B\nabla\theta|^2 \right\rangle ,  
\end{eqnarray} 
The last term on the LHS of each of Eqs.~\eq{corr} describes  {\em
spatial} flux of the corresponding quantity. In models of wall
bounded unstratified turbulence it is known that these terms are
very small almost everywhere. We do not have sufficient experience
with the stratified counterpart to be able to assert that the same
is true here. Nevertheless, for simplicity we are going to  neglect
these terms. It is possible to show that the accounting for these
terms does not influence much the results. Note that keeping these
terms turns the model into a set of differential equations which are
very cumbersome to analyze. This is a serious uncontrolled step in
our development, so we cross our fingers and proceed with caution.
Since these terms are neglected we do not provide here the explicit
expressions for $T_{ijk}$, $T_{ij}$, and $\B T$.

The first term on the RHS of the balance Eq.~\eq{corra} for the
Reynolds stresses is the ``Energy Production tensor" $\C P_{ij}$,
describing the production of the turbulent kinetic energy from the
kinetic energy of the mean flow, proportional to the gradient of the
mean velocity: 
\begin{subequations} \label{defs1} \begin{equation} \label{defs1A}
\C P_{ij} \equiv  - \tau_{ik}\, {\partial{U_j}}/{\partial{x_k}}
-\tau_{jk}\, {\partial{U_i}}/{\partial{x_k}} \ . \end{equation}
 The second term on the RHS of Eq.~\eq{corra}, $\C C_{ij} $, will be
 referred hereafter to as the
 ``Energy Conversion tensor". It
 describes the conversion of the turbulent  kinetic  energy into
 potential energy. This term is proportional to the buoyancy parameter
 $ \beta $ and the turbulent thermal flux $\B F$:
\begin{equation}\label{defs1B} \C C_{ij} \equiv
 - \beta\big (F_i\,\delta_{j\,z}+
F_j\,\delta_{i\,z}\big)\ .\end{equation}
The next term in the RHS of Eq.~\eq{corra} is known as the
``Pressure-rate-of-strain tensor": 
\begin{equation}\label{defs1C}  \C R_{ij} \equiv   \left\langle   {\widetilde p}
\,s_{ij}/{\rho\sb b}\right\rangle, \quad s_{ij}\equiv
 {\partial{u_i}}/{\partial{x_j}} + {\partial{
u_j}}/{\partial{x_i}} \ . \end{equation}
In incompressible turbulence its trace vanishes, therefore $ \C
R_{ij}$ does not contribute to the balance of the kinetic energy. As
we will show in Sec.~\ref{sss:PRS}, this tensor can be presented as
the sum of three contributions (Zeman, 1981),
\begin{equation}\label{rof}
\C R_{ij}= {R_{ij}\Sp{\, RI}} +{R_{ij}\Sp {\,IP}} +{R_{ij}\Sp{\,IC}}
\,,
\end{equation} 
in which ${R_{ij}\Sp{\, RI}}$ is  responsible for the nonlinear
process of isotropization of turbulence and is traditionally called
the ``Return-to-Isotropy", ${R_{ij}\Sp {\,IP}}$ is similar to the
energy production tensor~\eq{defs1A} and is called ``Isotropization
of Production". A new term, appearing in the stratified flow,
${R_{ij}\Sp{\,IC}}$,  is similar to the energy conversion
tensor~\eq{defs1B} and will be refereed to as the ``Isotropization
of Conversion".

Consider  the balance of the turbulent thermal flux $\B F$,
\Eq{corrb}. The first term in the RHS, $\BC A$, describes the source
of $\B F$ and, by analogy with the energy-production tensor, $\C P
_{ij}$, is called ``Thermal-flux production vector". Like $\C P
_{ij}$, \Eq{defs1B}, it has the contribution, $A_i^{^{SU}} $,
proportional to the mean velocity gradient:
 \begin{eqnarray}\label{defOfA}\nonumber
 {\C A}_i& \equiv \ & { A_i^{^{SU}}}
 + { A_i^{^{S\Theta}}} + { A_i^{^{E\theta}}} \,,\quad
  { A_i^{^{SU}}}  \equiv  -\B F \cdot \B \nabla\, U_i \,,
  \\
 { A_i^{^{S\Theta}}} & \equiv  & -\tau_{ij}\,
 {\partial \Theta}/{\partial x_j}\,,
\quad { A_i^{^{E\theta}}} \equiv  2\,\beta\,E_{\theta}\,\delta_{i\,z}\,,
\end{eqnarray}
 and two additional contributions, related to the temperature
gradient and to the ``temperature energy", $E_\theta$, and the
buoyancy parameter. One sees, that in contrary to the oversimplified
assumption~\eq{dtH}, the thermal flux in such a turbulent media cannot be
considered  as proportional to the temperature gradient. It has also
a contribution proportional to the velocity gradient and even to
the square of the temperature fluctuations. Moreover, the RHS of the
flux-balance \Eq{corrb} has an additional term, the
``Pressure-temperature-gradient vector" which, similarly to the
pressure-rate-of-strain tensor~\eq{rof}, can be divided into three
parts (Zeman, 1981): 
\begin{equation}\label{defs1f} 
\BC B  \equiv    \left\langle {\widetilde p }\, \B \nabla \theta /\rho_b
\right\rangle  = \B B \Sp {RD}  + \B B \Sp{SU} +  \B B \Sp{E\theta}\
.
\end{equation}
As we will show in Sec.~\ref{sss:PRS} the first contribution, $
B_i\Sp {RD}\propto  \left\langle  u\, u\, \nabla\!_i\, \theta \right\rangle $ is responsible
for the nonlinear flux of  $\B F$ in the space of scales toward
smaller scales, similarly to the correlation $ \left\langle  u\, u\, u \right\rangle $, which
is responsible for the flux of kinetic energy $ \left\langle  u^2 \right\rangle\!/2 $ toward
smaller scales. The correlation $ B_i\Sp {RD}
 \propto  \left\langle  u\, u\, \nabla\!_i\, \theta \right\rangle $ may be understood as the nonlinear contribution to the dissipation of the thermal flux. Correspondingly
we will call it  ``Renormalization of the Thermal-Flux Dissipation"
and will supply it with a superscript ``~$\Sp {RD}$~". The next  two terms  in the
decomposition~\eq{defs1f} are  ${B_i^{^{SU}}}\propto S\Sb U$ and
${B_i^{^{E\theta}}}\propto E_\theta$.  They describe the
renormalization of the thermal-flux production terms
${A_i^{^{SU}}}\propto S\Sb U$ and ${A_i^{^{E\theta}}}\propto
E_\theta$, accordingly. 
\end{subequations}
\subsection{\label{ss:cons}Conservation of total mechanical energy
in the exact balance equations}

The total mechanical energy of temperature stratified turbulent
flows consists of three parts with densities (per unit mass): $E=
E_{\C K}+ E\Sb K+E\Sb P$, where $E_{\C K}=|\B U|^2/2$ is the
density of kinetic energy of the mean flow, $E\Sb K=\tau_{ii}/2$
is the density of turbulent kinetic energy and $E\Sb P={\beta}
E_\theta/{S_\Theta}$ is the density of potential energy,
associated with turbulent density fluctuation $ \widetilde \rho= \widetilde \beta\,
\theta \rho_b$, caused by the (potential) temperature fluctuations
$\theta$, and $S_\Theta = d\,  {\Theta}/d z$.  

The balance Eq. for  $E_{\C K}$ follows   from \Eq{mean}:
\begin{subequations}\label{balE}
\begin{eqnarray}\label{balETm} 
 {D E_{\C K}}/{D\,  t}+ \nu  \left( \nabla_{\!\! j}  U_i \right) ^2+
 \nabla_{\!\! j}\, (U_i  \, \widetilde{\tau}_{ij}) = ~~~~~\nonumber \\%
 \left[\mathrm{source}\ E_{\C K}\right] + \tau_{ij}\nabla_{\!\! j} \, U_i\,,
\end{eqnarray}
with the help of identity: $ U_i\nabla_{\!\! j}\, \tau_{ij}\equiv
\nabla_{\!\! j}\, (U_i \tau_{ij})-\tau_{ij} \nabla_{\!\! j}\, U_i$
and definition~\eq{def-tau}. The terms on the LHS of this Eq.,
proportional to $\nu$ and $ \widetilde{\tau}_{ij}$ respectively, describe
the dissipation and the spatial flux of $ E_{\C K}$. The term
[source $E_{\C K} $]  on the RHS of \Eq{balETm} describes the
external source of energy, originating from the boundary conditions
described above, and $\tau_{ij}\nabla_{\!\! j} \, U_i$ describes the
kinetic energy out-flux from the mean flow to turbulent subsystem.

The balance Eq. for the turbulent kinetic energy follows directly
from~\Eq{corra}:  
\begin{equation}\label{balETt} 
 {D\, E\Sb K}/{D\,  t}+ \big[  \varepsilon _{ii}+\nabla_{\!\! j}\,
T_{iij}\big]/2=- \tau_{ij} \nabla_{\!\! j} \, U_i +\beta  F_z\ . 
\end{equation}
On the LHS of \Eq{balETt} one sees the dissipation and spatial flux
terms. The first term on the RHS originates from the energy
production, $\frac12 \, \C P_{ii}$, defined by \Eq{defs1A}. This
term has an opposite sign to the last term on the RHS of \Eq{balETm}
and describes the production of the turbulent kinetic energy from
the kinetic energy of the mean flow. The last term on the RHS of
\Eq{balETt} originates from the energy conversion tensor $\frac12 \,
\C C_{ii}$, \Eq{defs1B}, and describes the conversion of the
turbulent kinetic energy into potential one.

According to the last of Eqs.~\eq{corr}, one gets the balance
equation for the potential energy $E\Sb P$; multiplying \Eq{corrc}
for $E_\theta$ by $\beta /S_\Theta$:
\begin{equation}\label{balEP}
  {D\, E\Sb P}/{D\,  t}+  \beta \Big [ \epsilon +
 \nabla_{\!\! j} T_j \Big ]/S_\Theta  = -\beta F_z\ .\end{equation}
 \end{subequations}
 The RHS of this Eq. [coinciding up to a sign with the last term
  on the RHS of \Eq{balETt}] is the source of potential energy
(from the kinetic one).

In the sum of the three balance equations, the conversion terms (of the
kinetic energy from the mean to turbulent flows and of the turbulent
kinetic energy to the potential one) cancel and one gets the total mechanical
energy balance: 
\begin{equation}\label{bal-tot}
 {D\, E\ }/{D\,  t}+ [\mbox{diss }E] + \B \nabla\,
 [\mbox{flux} E]= [\mbox{source } E_{\C K}]\ .
 \end{equation}
This equation exactly respects the conservation of total mechanical
energy in the dissipation-less limit,  irrespective of the closure
approximations. This is because the energy production and conversion
terms are exact and do not require any closures, while the
pressure-rate-of-strain tensor, that requires some closure, does not
contribute to the total energy balance.

\section{\label{s:closure} The Closure Procedure and the resulting model}
In this section we describe the proposed closure procedure that
results in a model of  stably stratified TBL. In developing this
model we strongly rely  on the analogous well developed modeling of
standard (unstratified) TBL. The final justification of this
approach can be done only  in comparison to data from experiments
and simulations. We will do below what we can to use the existing
data, but we propose at this point that much more experimental and
simulational work is necessary to solidify all the steps taken in
this section.

\subsection{\label{sss:PRS} Pressure-Rate-of-Strain tensor $\C R_{ij}$ and Pressure-Temperature-Gradient vector $\BC B$} 
The correlation functions $\C R_{ij}$   and $\BC B$, defined by
Eqs.~\eq{defs1C} and \eq{defs1f}, include fluctuating part of the
pressure $\widetilde  p$.  The Poisson's equation for $\widetilde  p$
 follows from  \Eq{fluct}: $
\Delta \widetilde p = \rho \sb b\Big[ -\nabla _i \nabla _j  \left(  u _i u _j - \left\langle u
_i u _j \right\rangle
 +U_i u_j +U_j u_i \right)  +\beta\nabla_z\theta\Big]
$.
 The solution of this equation includes a harmonic part, $\Delta \widetilde p =
 0$, which is  responsible for sound propagation and does not
 contribute to  turbulent dynamics at small Mach numbers. Thus this
 contribution can be neglected. the inhomogeneous solution
 includes  three parts $\widetilde {p}=\rho \sb b[p_{uu}+p_{Uu}+p_\theta]$,
 where
  \begin{eqnarray}\label{pr1}
  p_{uu}&=&
 \Delta^{-1}\nabla _i \nabla _j \left(     \left\langle u _i u _j \right\rangle - u _i u _j \right) \,,\\ \nonumber
p_{Uu}&=& \Delta^{-1}\nabla _i \nabla _j \left(   U_iu_j+ U_ju_i \right) \,, \
p_\theta= \beta\Delta^{-1}\nabla_z\ \theta \ ,
\end{eqnarray}
and the inverse Laplace operator $ \Delta^{-1}$ is defined as usual
in terms of an integral over the Green's function.

Correspondingly the correlations  $\C R_{ij}$   and $\BC B$ consist
of three terms, Eqs.~\eq{rof} and \eq{defs1f}, in which 
 \begin{eqnarray}\label{decP}   &&R_{ij}\Sp{RI} = \left\langle
 p_{uu} s_{ij} \right\rangle\,, \       R_{ij}\Sp{IP}
 \equiv   \left\langle  p_{Uu}\, s_{ij}\right\rangle , \
R_{ij}\Sp{IC} \equiv  \left\langle  p_\theta s_{ij} \right\rangle,~\\
&&  \B B_i\Sp {RD}  = \left\langle  p _{uu} \B \nabla \theta  \right\rangle ,\ \B B ^{^{SU}}
   \equiv     \left\langle  p _{Uu} \B \nabla \theta  \right\rangle ,\ \B B_i^{^{E\theta}} \equiv
    \left\langle  p _\theta \B \nabla \theta  \right\rangle \nonumber \ .
   \end{eqnarray} 
All of those terms  originating from $p_{uu}$ are the most
problematic because they introduce coupling to triple correlation
functions: ${R_{ij}\Sp{\, RI}}\propto  \left\langle u_iu_ju_k \right\rangle $ and  $\B
B\Sp{RD}\propto  \left\langle  u^2 \B \nabla \theta  \right\rangle $. Thus they require
closure procedures whose justification can be only tested
a-posteriori against the data.

Having in mind to simplify the model in most possible manner, we
adopt for the diagonal part of the Return-to-Isotropy tensor, the
simplest Rota form (Rotta, 1951)
\begin{subequations}\label{ROF}
\begin{equation}\label{RIa} 
{R_{ii}\Sp{\, RI}} \simeq   -\gamma \Sb {RI} \left(   \tau_{ii} - 2\,E\Sb
K / 3 \right)  \,, 
\end{equation} 
in which  $\gamma \Sb{RI}$ is the relaxation frequency of diagonal
components of the Reynolds-stress  tensor toward its isotropic form,
$2 E\Sb K/3$.  The parametrization of  $\gamma \Sb{RI}$    will be
discussed later.  The tensor ${R_{ij}\Sp{\, RI}}$ is traceless,
therefore the frequency $\gamma \Sb{RI}$ must be the same for all the
diagonal components of ${R_{ii}\Sp{\, RI}}$. On the other hand there
are no reasons to assume that off-diagonal terms have the same
relaxation frequency.   Therefore, following~L'vov~et~al.~(2006a) we
assume that 
\begin{equation}\label{RIb} 
{R_{ij}\Sp{\, RI}} \simeq   -   \widetilde \gamma \Sb {RI} \tau_{ij}  \,, \quad
i\ne j\,,
\end{equation}
with, generally speaking, $\widetilde \gamma \Sb {RI}\ne \gamma \Sb {RI}$. Moreover, on the
intuitive level,  we  can expect that off-diagonal terms should
decay faster then the diagonal ones, i.e. $\widetilde \gamma \Sb {RI}> \gamma \Sb
{RI}$. Indeed, our analysis of DNS results  shows
that $\widetilde \gamma \Sb {RI}/\gamma \Sb {RI}\simeq 1.46$ (L'vov~et~al., 2006b).

The term $\B B\Sp{RD}$ also describes return-to-isotropy due to
nonlinear turbulence self interactions  (Zeman, 1981), and may be
modeled as: %
\begin{equation}\label{RD1} {B_i\Sp {RD}} =-\gamma  \Sb {RD} F_i\ .
\end{equation} 
 This equation dictates the vectorial structure of $
{B_i\Sp {RD}} \propto F_i$, which will be confirmed below. The rest
can be understood as the definition of the $\gamma  \Sb {RD}$ as  the
relaxation frequency of the thermal flux. Its parametrization  is
the subject of further discussion in Sec. \ref{ss:closure}.

The traceless ``Isotropization-of-Production" tensor, ${R_{ij}\Sp
{\,IP}}$, has a very similar structure to the production tensor, $\C
P_{ij}$, \Eq{defs1A},  and thus is traditionally  modeled in terms of
$\C P_{ij}$~(Pope, 2001):
\begin{equation} \label{IP}
{R_{ij}\Sp {\,IP}} \simeq  -C\Sb {IP} \left(    \C \mathcal{P}_{ij} - \delta_{ij}\,{\C P}/3
 \right) \,,  \quad \C P\equiv   \mbox{Tr} \,\{ \C \mathcal{P}_{ij}\}\ .
\end{equation}
The accepted value of the numerical constant $ C\Sb {IP} = \frac35\ $
(Pope, 2001).

The traceless ``Isotropization-of-Conversion" tensor, ${R_{ij}\Sp
{\,IC}}$ does not exist in unstratified TBL. Its structure is very
similar to the conversion tensor, $\C C_{ij}$, \Eq{defs1B}. Therefore
it is reasonable to  model it in the same way in terms of $\C
C_{ij}$ (Zeman, 1981): \begin{equation} \label{IC}
{R_{ij}\Sp {\,IC}} \simeq  -C\Sb {IC} \left(
  \C C_{ij} -\delta_{ij}\, {\C C}/3
 \right) \,,  \quad \C C\equiv   \mbox{Tr} \,\{ \C C_{ij}\}\,,
\end{equation}
with some new constant $ C\Sb {IC}$ . 

The renormalization of production terms ${B_i^{^{SU}}}$ and
${B_i^{^{E\theta}}}$ are very similar to the corresponding thermal flux
production terms, ${A_i^{^{SU}}}$ and ${A_i^{^{E\theta}}}$, defined
by Eqs.~\eq{defOfA}. Therefore, in the spirit of Eqs.~\eq{IP} and
\eq{IC}, they are modeled as follows:
\begin{eqnarray} 
{B_i^{^{SU}}}&=& (C_{_{SU}}-1){A_i^{^{SU}}}= (1-C_{_{SU}})
(\B F \cdot \B \nabla\,) U_i\,, \\ 
{B_i^{^{E\theta}}} &=& -(C_{_{E\theta}}+1) {A_i^{^{E\theta}}} =
 -2\,\beta\,(C_{_{E\theta}}+1)E_{\theta}\,\delta_{i\,z}\ .~~~~~~~~ 
\end{eqnarray} 
\end{subequations} 
Using this and \eq{pr1} one finds the sign of $C_{_{E\theta}}$:
 \begin{eqnarray}\nonumber
&& -\beta \left( C_{_{E\theta}} +1 \right)  E_{\theta} =  \left\langle \widetilde p_{\theta} \nabla_z
\theta \right\rangle  = \beta\langle (\nabla_z\theta) \Delta^{-1}
(\nabla_z\theta)\rangle,~~~\\ 
&& \label{CT}
C_{_{E\theta}} = - \left( 1 + {\langle (\nabla_z\theta) \Delta^{-1}
(\nabla_z\theta)\rangle}/{\langle \theta^2\rangle} \right)  < 0\ .
\end{eqnarray}
To estimate $C_{_{E\theta}}$ we assume that on the gradient scales
the temperature fluctuations are roughly isotropic,
and therefore we can estimate
 $\Delta= \nabla_x^2+\nabla_y^2+\nabla_z^2 \approx 3\nabla_z^2$.
  Introducing this
estimate and integrating by parts leads to $ C_{_{E\theta}}\approx
-2/3$.

\subsection{\label{sss:dis}Reynolds-stress-, thermal-flux-, and
 thermal-dissipation} 
Far away from the wall and for large Reynolds numbers the
dissipation tensors are dominated by the  viscous scale motions,
at which turbulence can be considered as isotropic. Therefore, the
vector $\B \epsilon$ should vanish, while the tensor $\varepsilon _{ij}$,
\Eq{diss}, should be diagonal:  
\begin{subequations}\label{distau1}
\begin{equation}\label{distau1a} \epsilon_i=0\,, \quad \varepsilon _{ij}=
2\, \gamma _{uu} \,E\Sb K \, \delta _{ij}/3\,, 
\end{equation}
where  the numerical prefactor $\frac23 $ is chosen such that
$\gamma _{uu}$ becomes the relaxation frequency of the turbulent kinetic
energy. Under stationary conditions the rate of turbulent kinetic
energy dissipation is equal to the energy flux through scales, that
can be estimated as $ \left\langle  u u u  \right\rangle / \ell$, where $\ell$ is the outer
scale of turbulence. Therefore, the natural estimate of $\gamma _{uu}$
involves the triple-velocity correlator, $ \gamma _{uu}\sim  \left(   \left\langle  u u u
 \right\rangle / \ell  \left\langle  uu  \right\rangle  \right) $, 
exactly in the same manner, as the Return-to-Isotropy frequencies,
$\gamma   \Sb{RI}$ and $ \widetilde {\gamma }\Sb{RI}$ in Eqs.~\eq{RIa} and \eq{RIb}.
Similarly,
\begin{equation}\label{distau1b}
\varepsilon  =\gamma  _{\theta\theta}\, E_\theta  \,,    \quad \gamma _
{\theta\theta} \sim   \left\langle \theta \theta u  \right\rangle \big / \ell  \left\langle \theta
\theta \right\rangle  \ . \end{equation}\end{subequations} 
\subsection{\label{ss:balance}Stationary balance equations
 in plain geometry}

In the plane geometry, the equations simplify further. The mean
velocity is oriented in the (streamwise) $\widehat{\bf{x}} $ direction and all
mean values depend  on the vertical (wall-normal) coordinate $z$
only: 
$\B U  =  U (z)\,\widehat{\bf{x}}  $, $ {\Theta}= {\Theta}(z)$, $
\tau_{ij}=\tau_{ij}(z)$, $ ~\B F=\B F(z)$, $E_\theta  =E_\theta (z)$.
Therefore     $ \left( \B U\cdot\B \nabla \right)  \left\langle  \dots \right\rangle  = 0$, and in the
stationary case, when $\partial \ /\partial  t= 0$,  the mean convective
derivative  vanishes: $D\ /D\, t=0$.  Moreover due to the $y\to -y$
symmetry of the problem the following correlations vanish:
 $
\widetilde \tau_{xy}=\widetilde \tau_{yz}=\widetilde F_y=0$.
 The only non-zero components of the mean velocity and
temperature gradients are:
 \begin{equation}\label{Shears} S_{_U}\equiv  {d U}/{d z}\,, \quad
S_{_\Theta}\equiv   {d \Theta}/{d z}\ . \end{equation} 

\subsubsection{Equations for the mean velocity and temperature profiles}
 Having in mind Eqs. of Sec.~\ref{ss:balance} and integrating    Eqs.
\eq{mean} for $U_x$ and $\Theta$   over $z$,   one gets equations
for the total (turbulent and molecular) mechanical-momentum flux,
$\widetilde \tau (z)$, and thermal flux, $\widetilde F$, toward the wall 
\begin{subequations} \label{TBL-Sim}
  \begin{eqnarray}  \label{Sim-U-x}
 \widetilde \tau_{xz}(z)=-\nu\,S_{_U} +\tau_{xz}   \Rightarrow
 \widetilde \tau_{xz}(0)\equiv - \tau_* \,, \\
\label{Sim-Teta-Avg}
  \widetilde  F_z(z)=-\chi\,S_{_\Theta} + F_z  \Rightarrow \widetilde F_z(0)\equiv  - F_*\ . 
\end{eqnarray}\end{subequations} 
The total flux of the $x$-component of the
mechanical moment in $z$-direction is 
$ \rho_b\widetilde \tau_{xz} (z)  \equiv    \int dz ({\partial  \left\langle  p \right\rangle /\partial  x}) +
\mbox{const}$.  Generally speaking, $\widetilde \tau_{xz} (z)$ depends on
$z$. For example, for the pressure driven planar channel flow (of
the half-wight $\delta$) $\rho_b\widetilde \tau_{xz} (z)   =
 (\partial  \left\langle  p \right\rangle /\partial  x)(\delta-z)<0$.

Relatively close to the ground, where $z\ll \delta$, the $z$
dependence of $\widetilde \tau_{xz}(z)$ can be neglected. In the absence of
the mean horizontal pressure drop and spatial distributed heat
sources $\widetilde \tau$ and $\widetilde F$ are $z$-independent, and thus equal to
their values at zero elevation, as indicated in Eqs.~\eq{TBL-Sim}
after ``$\Rightarrow$"-sign. Notice, that in our case of stable
stratification both vertical fluxes, the $x$-component of the
mechanical momentum, $\widetilde \tau_{xz}$, and the thermal flux, $\widetilde F_z$,
are directed toward the ground, i.e. negative. For the sake of
convenience, we introduce in Eqs.~\eq{TBL-Sim} notations for their
(positive) zero level absolute value: $\tau_*$ and $F_*$.

Recall that in the plain geometry $U_z=0$. Nevertheless one can
write an equation for $U_z$:  
\begin{equation} \label{Sim-U-z}
 d  \left( \tau_{zz} +{ \left\langle  p  \right\rangle /\rho_b} \right) /d z   = \beta\,\Theta \,,
\end{equation}
 which describes a turbulent correction ($\propto \tau_{zz}$) to the
 hydrostatic equilibrium. Actually, this equation determines
 the profile of $ \left\langle  p \right\rangle $, that does not appear in the system of
balance equations ~\eq{TBL-Sim}.
\subsubsection{Equations for the pair (cross)-correlation functions}
  Consider first
  the balance Eqs.~\eq{corra} for the
diagonal components of the Reynolds-stress tensor in algebraic
model (which arises when we neglect the spatial fluxes): 
\begin{eqnarray}\nonumber
&&   \Gamma  E\Sb K +3 \gamma\Sb{RI} \tau_{xx} /2 = - \Big(3-
2\,C\Sb{IP} \Big)\tau_{xz}S_{_U} - C\Sb{IC}\,\beta \,F_{z}\,,~~~~ \\
\label{taup}  
&& \Gamma E\Sb K +3\gamma\Sb{RI} \tau_{yy}/2   = -
C\Sb{IP}\tau_{xz}S_{_U} - \,C\Sb{IC}\,\beta \,F_{z}\,,\\ \nonumber
 &&  \Gamma E\Sb K + 3\gamma\Sb{RI} \tau_{zz} /2 =
-C\Sb{IP}\tau_{xz}S_{_U} +\Big(3 +2\,C\Sb{IC} \Big)\beta \,F_{z}\
.
\end{eqnarray}
where $\Gamma\equiv \gamma _{uu}-\gamma  \Sb{RI}$.
 The LHS of these equations includes the dissipation and
Return-to-isotropy terms. On the RHS  we have the    kinetic energy
production and isotropization of production terms (both proportional to
$S_{_U}$) together with the conversion and isotropization of
conversion terms, that are proportional to the vertical thermal flux
$F_z$. The horizontal component of the thermal flux,  $F_x$, does
not appear in these equations.

System~\eq{taup} allows to find anisotropy of the
turbulent-velocity fluctuations and to get  the balance Eqs. for the
turbulent kinetic energy with the energy production and conversion
terms on the RHS: 
 \begin{subequations}\label{tau1} \begin{eqnarray}
3\tau_{xx} &=& 2 \big\{ [2(1 -C\Sb{IP})\Gamma_{uu}/\gamma\Sb{RI} +1
] E\Sb K \\ \nonumber  &&  -(3 -2C\Sb{IP} +C\Sb{IC})\, \beta F_z /
\gamma\Sb{RI} \big\} \,,
\\ 
3\tau_{yy} &=& 2 \big\{ [(C\Sb{IP}-1)\Gamma_{uu}/\gamma\Sb{RI}  +1 ]
E\Sb K \\ \nonumber  &&
-(C\Sb{IP} +C\Sb{IC}) \beta F_z/\gamma\Sb{RI}  \big\} \,, \\
3\tau_{zz} &=& 2 \big\{ [ \left( C\Sb{IP}-1 \right)
 \Gamma_{uu}/\gamma\Sb{RI}  +1 ]E\Sb K \\ \nonumber  &&   -
(C\Sb{IP} -2C\Sb{IC} -3 ) \, \beta F_z/\gamma\Sb{RI}\big\} \,, 
 \\ \label{bale}
 \Gamma_{uu}E\Sb K &=&   -\tau_{xz}S_{_U}+ \beta F_z \,,
\end{eqnarray} 
Equation~\eq{bale} includes   the only non-vanishing tangential
(off-diagonal) Reynolds stress $\tau_{xz}$ and has to be accompanied
with an equation for this object:
  \begin{equation}
\label{txz}
\widetilde  \gamma \Sb {RI} \tau_{xz}  =  \big(C\Sb{IP} -1\big)\tau_{zz}\,S_{_U}
+\big(1 +C\Sb{IC}\big)\beta\,F_x \ . \end{equation} 
\end{subequations}
\begin{subequations}
This equation manifests  that the tangential Reynolds stress
$\tau_{xz}$, that determines the energy production [according to
\Eq{bale}], influences, in its turn, on the value of the
streamwise thermal flux $F_x$,  which therefore  effects on the
turbulent kinetic energy production.

 As we mentioned, in the plain geometry $\widetilde  F_y=0$.
  Equations~\eq{corrb} for the  $F_x$ and $F_z$  in this case take the form: 
\begin{eqnarray}  \label{Sim-F-x}
\gamma\Sb{RD} F_x &=& - \left( \tau_{xz} S_{_\Theta} +C\Sb{SU} F_z
S_{_U} \right) \,, \\ 
\label{Sim-F-z}
\gamma\Sb{RD} F_z &=& - \left( \tau_{zz} S_{_\Theta} +2\,C\Sb{E\Theta}\,
\beta E_{\theta} \right) \,,  
\end{eqnarray} 
in which the RHS describes the thermal-flux production, corrected by
the isotropization of production terms.

The last   \Eq{corrc} for $E_{\theta}$,   represents the balance
between the dissipation (LHS) and  production (RHS):
\begin{equation} \label{Sim-ET}
 \gamma_{\theta\theta}\,E_\theta = -F_z\,S_{_\Theta}\ .
\end{equation}
\end{subequations} 

\subsection{\label{ss:closure} Simple closure of time-scales and
the balance equations in the turbulent region}

At this point we follow a tradition in modeling of all the nonlinear
inverse time-scales by dimensional estimates (Kolmogorov, 1941): 
\begin{eqnarray} \label{freqs}
  \gamma _{uu}&=&c_{uu} \sqrt{E\Sb K} \big / \ell\,, \quad 
  \gamma \Sb {RI}= C\Sb{RI} \gamma _{uu}\,,  \\ \nonumber 
  \widetilde  \gamma\Sb {RI}& = &  \widetilde  C\Sb {RI} \gamma\Sb {RI}\,,\quad 
  \gamma_{\theta\theta}  = C_{\theta\theta} \gamma _{uu}\,, \quad
  \gamma\Sb{RD}= C_{u\theta}\gamma _{uu}\ .
\end{eqnarray}
Remember that  $\ell $ is the ``outer  scale of turbulence". This
scale equals to $z$ for $z< L$, where $L$ is the Obukhov
length (definition is found below).

Detailed analysis of experimental, DNS and LES data (see
L'vov~et~al.,~2006, and references therein) shows that for
unstratified flows, $\textrm{g}=0$, the anisotropic boundary
layers exhibits values of the Reynolds stress tensor that can be
well approximated by the values $\tau_{xx}=E\Sb K$,
$\tau_{yy}=\tau_{zz}=E\Sb K/2$. In our approach this dictates the
choice $C\Sb{RI} = 4(1-C\Sb{IP})$. Also we can expect that
$\tau_{yy}$ is almost not affected by buoyancy. This gives simply
$ C\Sb{IC} = -C\Sb{IP}$.  If so, Eqs.~\eq{tau1} with the
parametrization~\eq{freqs} can be identically rewritten as
follows: 
\begin{subequations}\label{simple}
\begin{eqnarray} \nonumber 
&&   \tau_{xx} =   E\Sb K -\frac{\beta F_z}{2\,\gamma_{uu}}\,,
\quad
\tau_{yy}  = \frac{E\Sb K}2\,,\quad  
\tau_{zz}  = \frac{E\Sb K}2 +\frac{\beta F_z}{2\,\gamma_{uu}}\,, \\
\label{simpleA}
 &&  \gamma_{uu}E\Sb K = \beta F_z -\tau_{xz}S_{_U} \,, \quad
\gamma _{uu}=c_{uu} \sqrt{E\Sb K} \big / \ell \,, \\ \nonumber &&  ~~~~~~~~~~~
4\,\widetilde C\Sb{RI}\,\gamma_{uu} \tau_{xz} = \beta\,F_x -\tau_{zz}\,S_{_U}
\  .
\end{eqnarray}  
For  completeness we also repeated here the
parametrization~\eq{freqs} of $\gamma _{uu}$. Finally we present the
version of the balance Eqs. for the thermal flux \eq{Sim-F-x},
\eq{Sim-F-z}, and for the ``temperature energy", \eq{Sim-ET}, after
all the simplified assumptions:
\begin{eqnarray} \nonumber
C_{\theta\theta}\,\gamma_{uu}E_\theta &=& -F_z\,S_{_\Theta}\,, \\
\label{simpleB} 
C_{u\theta}\,\gamma_{uu} F_x &=& - \left( \tau_{xz}S_{_\Theta}
 +C\Sb{SU}F_z\, S_{_U} \right) \,,  \\ \nonumber
C_{u\theta}\,\gamma_{uu} F_z &=& - \left( \tau_{zz}  S_{_\Theta}
 +2\,C\Sb{E\Theta}\,\beta  E_{\theta} \right) \ .
\end{eqnarray} \end{subequations}  

\subsection{\label{ss:wall-units}Generalized wall normalization}
The analysis of the balance Eqs.~\eq{simple}   is drastically
simplified if they are presented in a dimensionless form.
Traditionally, the conventional ``wall units" are introduced via the
wall friction velocity $u_*  \equiv   \sqrt{ \tau_*}$, and the viscous length-scale
$ \lambda_* \equiv    {\nu}/{u_*}$.
A wall unit for the temperature $\theta_*\equiv  {F_*}/{u_*}$ is defined via
the thermal flux at the wall  and friction velocity. Subsequently,
$ \B r^+ \equiv  { \B r}/{\lambda_*}$,     $t^+ \equiv   {t\, \lambda_*}/{
u_*}$, $\BC U^+ \equiv    {\BC U}/{ u_*}$, $p^+ \equiv  p/\rho_b\,u_*^2$, $
 {\Theta}^+   \equiv    { {\Theta}}/{\theta_*}$, $
\theta^+ \equiv   {\theta}/{\theta_*}$,  etc. Then the governing
Eqs.~\eq{GovEq} take the form:
\begin{eqnarray}\nonumber 
 {\C D^+ \, \BC U^+}/{\C D t ^+ } + \B {\nabla}^+ p^+  &=& 
 { \widehat{\bf{z}} }\,\Theta^+\sb {\,d}/{L^+} \ +\Delta^+\, \BC U^+ \,, \\
 \label{NSE-dim}
 {\C D^+ \,   \Theta^+\sb {\,d}}/{\C D t ^+ } &=&  
  \Delta^+\, \Theta^+\sb {\,d}/{\textrm{Pr}} \ .
\end{eqnarray}  
These Eqs. include two dimensionless parameters: the conventional
Prandtl number Pr$=\nu\big / \kappa$, and $L^+$ -- the
Obukhov length $L$ measured in wall units: $L\equiv  u_*^3\big /
\beta F_*$, $L^+\equiv L\big / \lambda_*$.
 We used here the modern
definition of the Obukhov length, which differs from the old
one by the absence of the \vK constant $\kappa$
in its denominator (Monin and Obukhov, 1954).

 Outside of the viscous sub-layer, where the  kinematic viscosity
and kinematic thermal conductivity can be ignored,  $L^+$ is the
only dimensionless parameter in the problem, which separates the
region of weak stratification, $z^+< L^+$, and the region of strong
stratification, where $z^+>L^+$.

Given the generalized wall normalization we introduce objects
with a superscript  ``~$^+$"
 in the usual manner:
 \begin{eqnarray}\nonumber
  S_{_U}^+ &\equiv &  t_*\, S_{_U}\,, \
S_{_\Theta}^+\equiv  {\lambda_* \, S_{_\Theta}}/{\theta_*}\,, \
\gamma^+\equiv  t_* \gamma\,, \ \tau_{ij}^+ \equiv    {\tau_{ij}}/{u_*^2}\,,\\
 \label{plus-m}\B F^+ &\equiv &
 {\B F}/{u_* \theta_*}\,,\ E_\theta^+ \equiv
 {E_\theta}/{\theta_*^2}\ .\end{eqnarray}
In the turbulent region, governed by $L^+$ only,   Eqs.~\eq{TBL-Sim}
simplify to $ \tau^+_{xz}  = -1$, $F^+_z = -1$. 
\subsection{\label{sss:resc} Rescaling symmetry and
$\ddag$-representation} Outside of the viscous region, where
Eqs.~\eq{simple} were derived, the problem has only one
characteristic length, i.e. the Obukhov scale  $ L$.
Correspondingly, one expects that the only dimensionless parameter
that governs the turbulent statistics in this region should be the
ratio of the outer scale of turbulence, $\ell$, to the Obukhov
length-scale $L$, which we denote as $ {\ell^\ddag}\equiv
 {\ell}/{L}=   {\ell^+} / L^+  $.
Indeed, introducing ``$\ddag$-objects": 
\begin{equation}\label{ddagb} 
 {\ell^\ddag}\equiv
 {\ell}/{L}\,, \quad
 S_{_U}^\ddag\equiv  S_{_U}^+{\ell^+}\,, \  S_{_\Theta}^\ddag\equiv
S_{_\Theta}^+{\ell^+}\,,  \end{equation} 
and using Eqs.~\eq{plus-m} one rewrites the balance Eqs.~\eq{simple}
as follows: 
{
\begin{subequations} \label{MM}
\begin{eqnarray} \label{9-eqA}
&& \hskip -1.3 cm \tau^+_{xx} =  E^+\Sb K + {{\ell^\ddag}  } /2
c_{uu}
\sqrt{E^+\Sb K}\,, \quad \tau^+_{yy}  =  {E^+\Sb K}/2\,,\\
\label{9-eqB}
&&  \hskip  -1.3 cm 2\, \tau^+_{zz}= {E^+\Sb K}  -{{\ell^\ddag}
 } / c_{uu} \sqrt{E^+\Sb K}\,,~~~~ \\ \label{9-eqC}
 &&  \hskip  -1.3 cm c_{uu}  { E^+\Sb K}^{3/2} =  {\ell^\ddag} F^+_z -\tau^+_{xz}
S^\ddag_{_U} \,, \\ \label{9-eqD}
\label{not-used}
&&  \hskip  -1.3 cm 4\,\widetilde C\Sb{RI}\,c_{uu} \sqrt{E^+\Sb K}
\tau^+_{xz} ={\ell^\ddag} F^+_x -\tau^+_{zz}\,S^\ddag_{_U} \,, \\
\label{9-eqE}
&&  \hskip  -1.3 cm  C_{\theta\theta}\,c_{uu} \sqrt{E^+\Sb K}
E^+_\theta = -F^+_z\,S^\ddag_{_\Theta}\,, \\ \label{9-eqF}
&&  \hskip  -1.3 cm C_{u\theta}\,c_{uu} \sqrt{E^+\Sb K} F^+_x = -
\tau^+_{xz}S^\ddag_{_\Theta} -C\Sb{SU}F^+_z\, S^\ddag_{_U} \,,
\\ \label{9-eqG}
&&  \hskip  -1.3 cm  C_{u\theta}\,c_{uu} \sqrt{E^+\Sb K} F^+_z = -
\tau^+_{zz}S^\ddag_{_\Theta} - 2\,C\Sb{E\Theta}\,{\ell^\ddag}
E^+_{\theta} \,,
\end{eqnarray}\end{subequations}}
These equations are the main result of  current
Sec.~\ref{s:closure}. It may be considered as ``Minimal Model" for
stably stratified TBL, that respects the conservation of energy,
describes anisotropy of turbulence and  all relevant fluxes
explicitly and, nevertheless is still simple enough to allow
comprehensive analytical analysis, that results in an approximate
analytical solution (with reasonable accuracy) for the mean
velocity and temperature gradients $S\Sb U$ and $S\Sb \Theta$,
and all second-order (cross)-correlation functions.

As expected, the only parameter appearing in the Minimal
Model~\eq{MM}   is  ${\ell^\ddag}$. The outer scale of turbulence,
$\ell$, does not appear by itself, only via the definition of
${\ell^\ddag}$ ~\eq{ddagb}. Therefore our goal now is to solve
Eqs.~\eq{MM} in order to find five functions of only one argument $
{\ell^\ddag} $: $S^\ddag _{_U}$, $S_{_\Theta}^\ddag$, $E^+\Sb K$,
$E_{\theta}^+$ and $ F^+_x$. After that we can specify the
dependence ${\ell^+}(z^+)$ and then reconstruct the $z^+$-dependence
of these five objects. 

\section{\label{s:res}  Results and discussion}

\subsection{\label{ss:strat41} Analytical solution of the Minimal-Model
balance equations~\eq{MM}}
\begin{figure*}
\begin{center}
 \includegraphics[width=0.315  \textwidth]{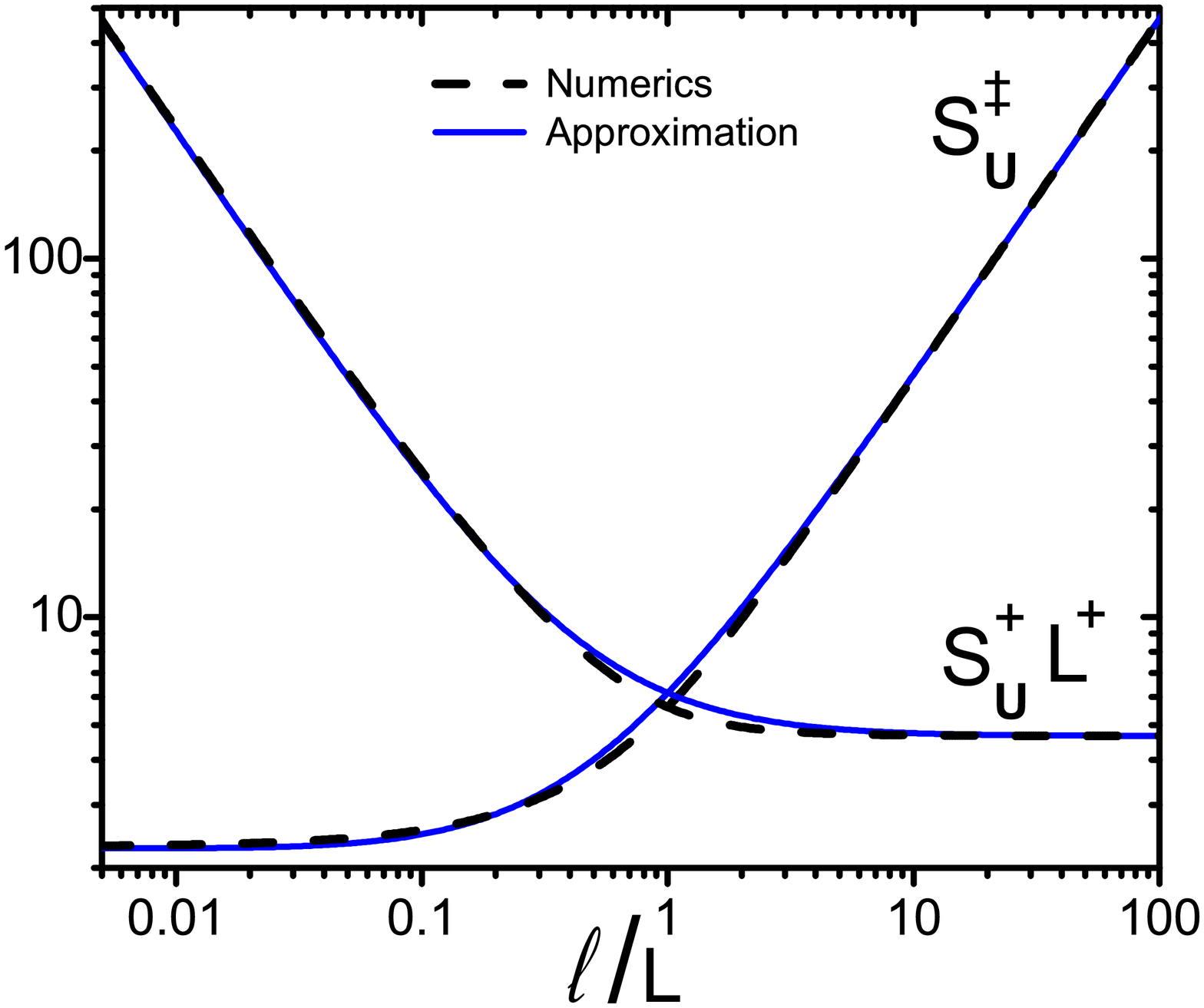}
 \includegraphics[width=0.316  \textwidth]{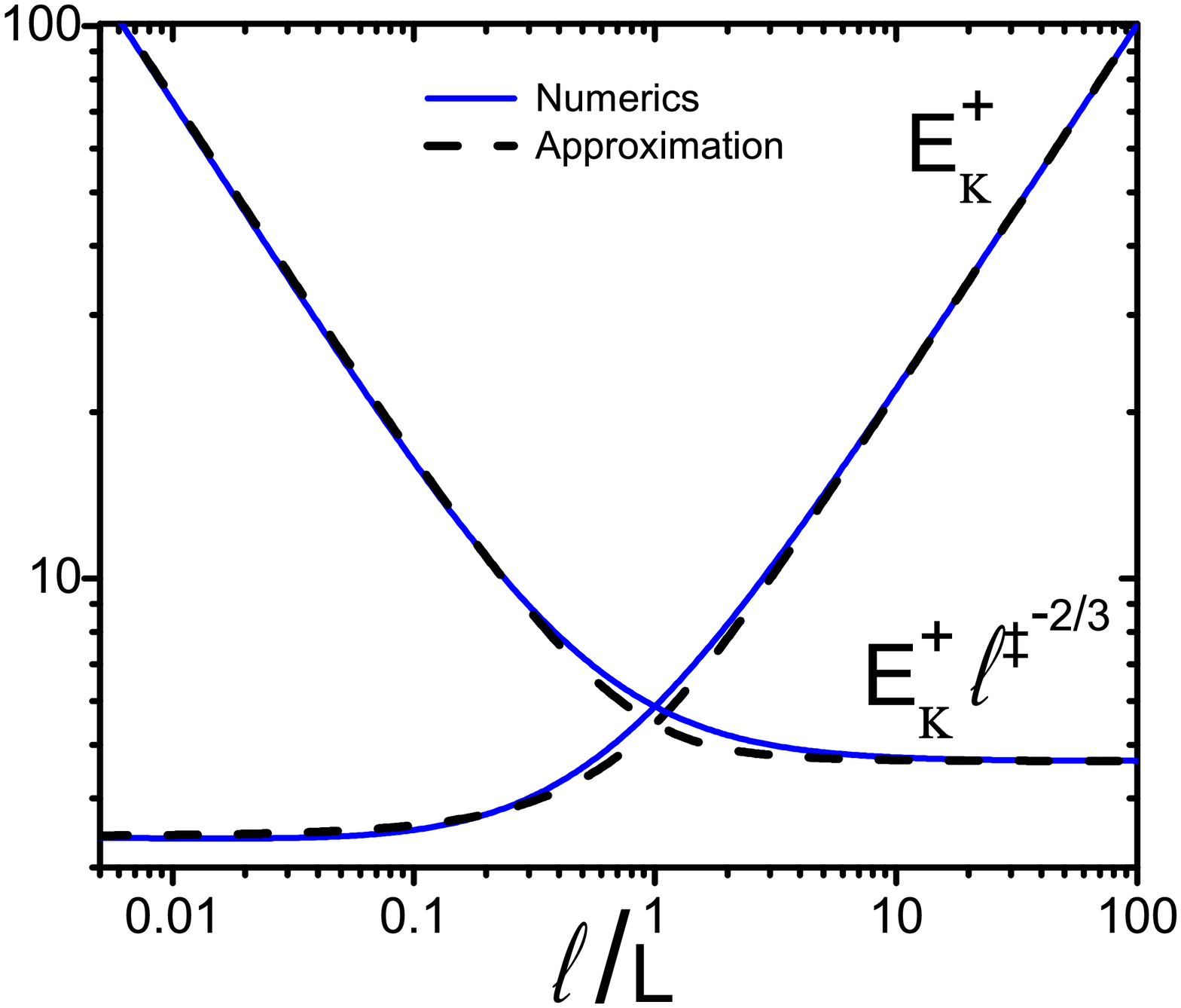}
 \includegraphics[width=0.315  \textwidth]{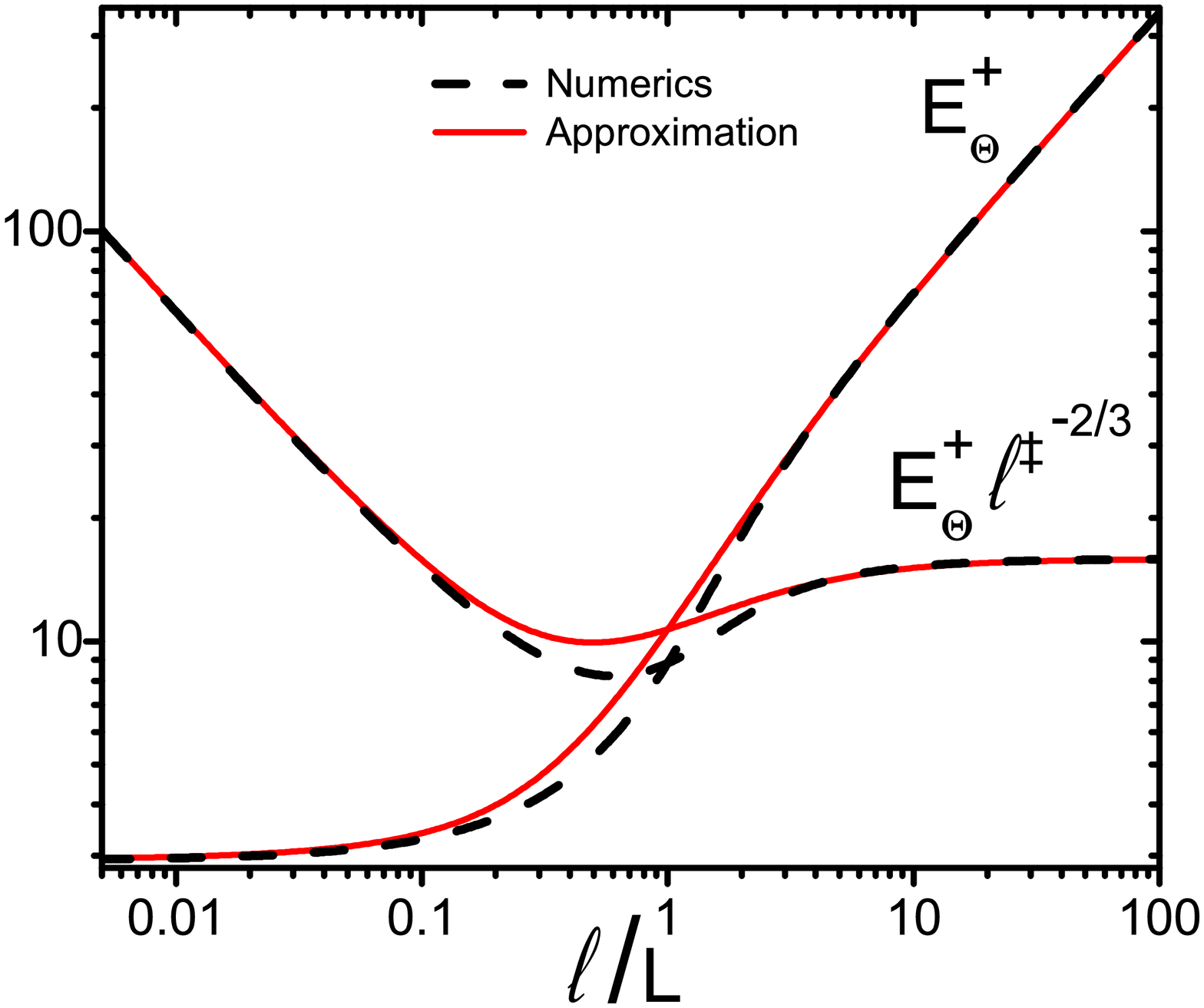}\\
 \includegraphics[width=0.317  \textwidth]{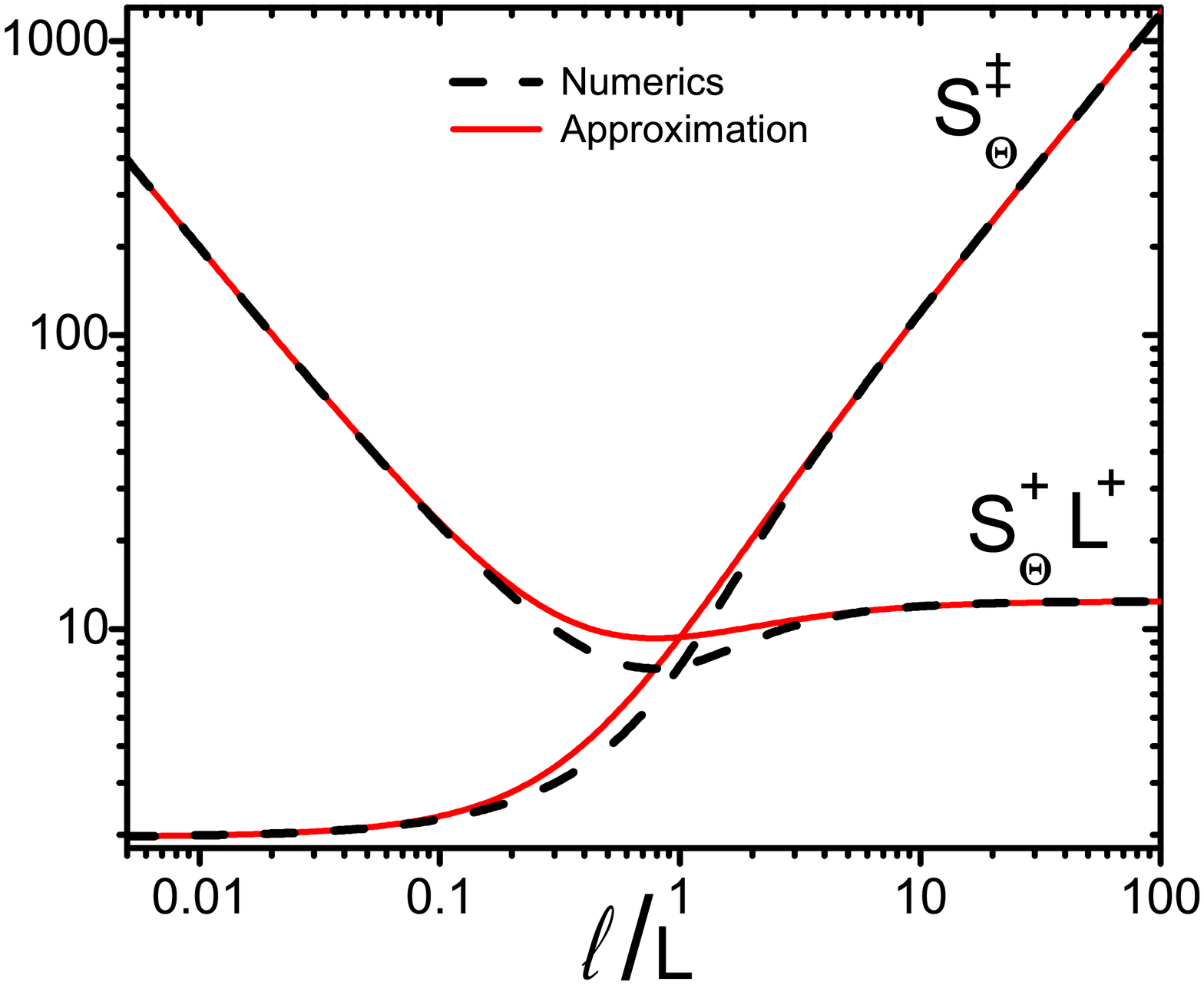}
 \includegraphics[width=0.312  \textwidth]{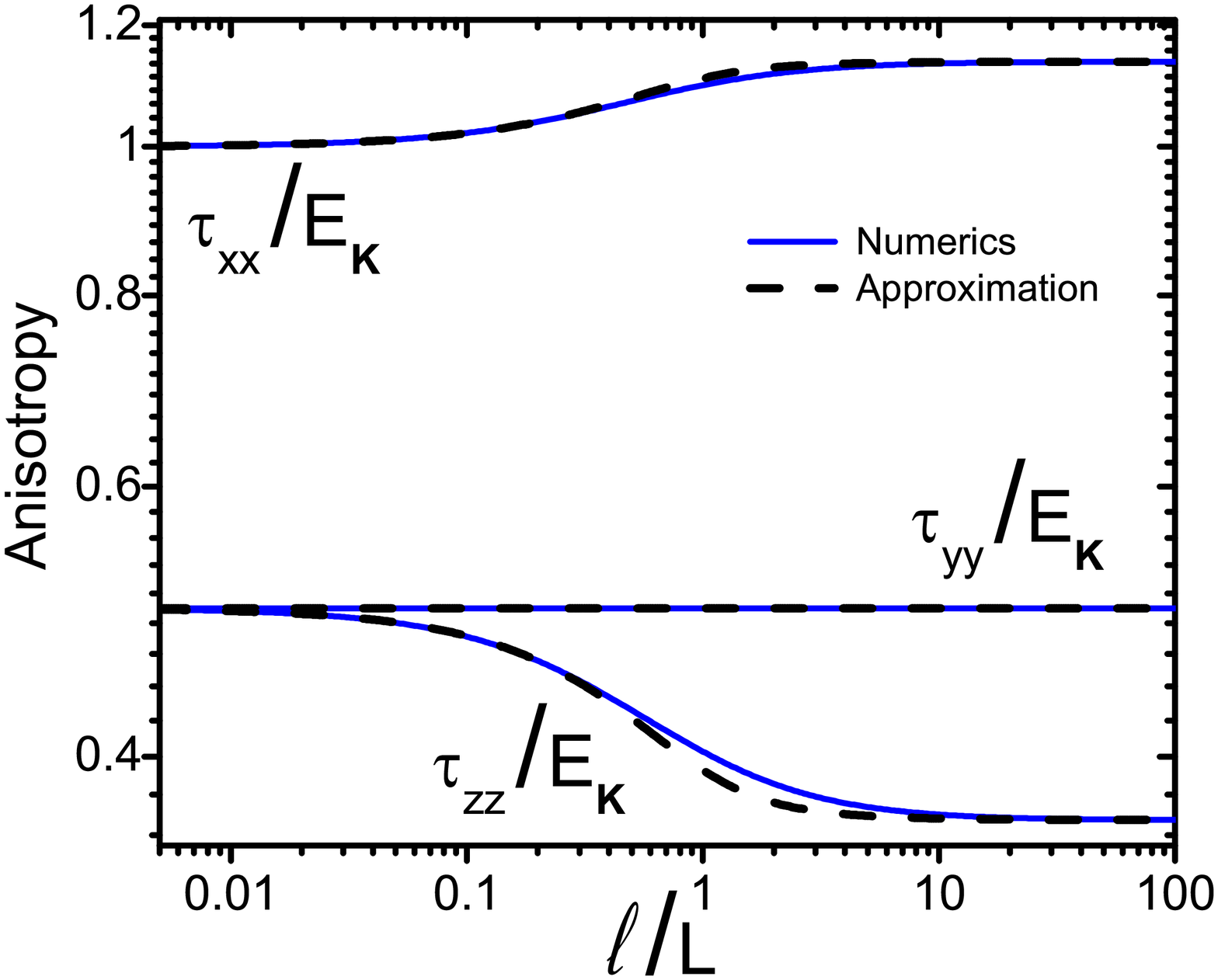}
 \includegraphics[width=0.315  \textwidth]{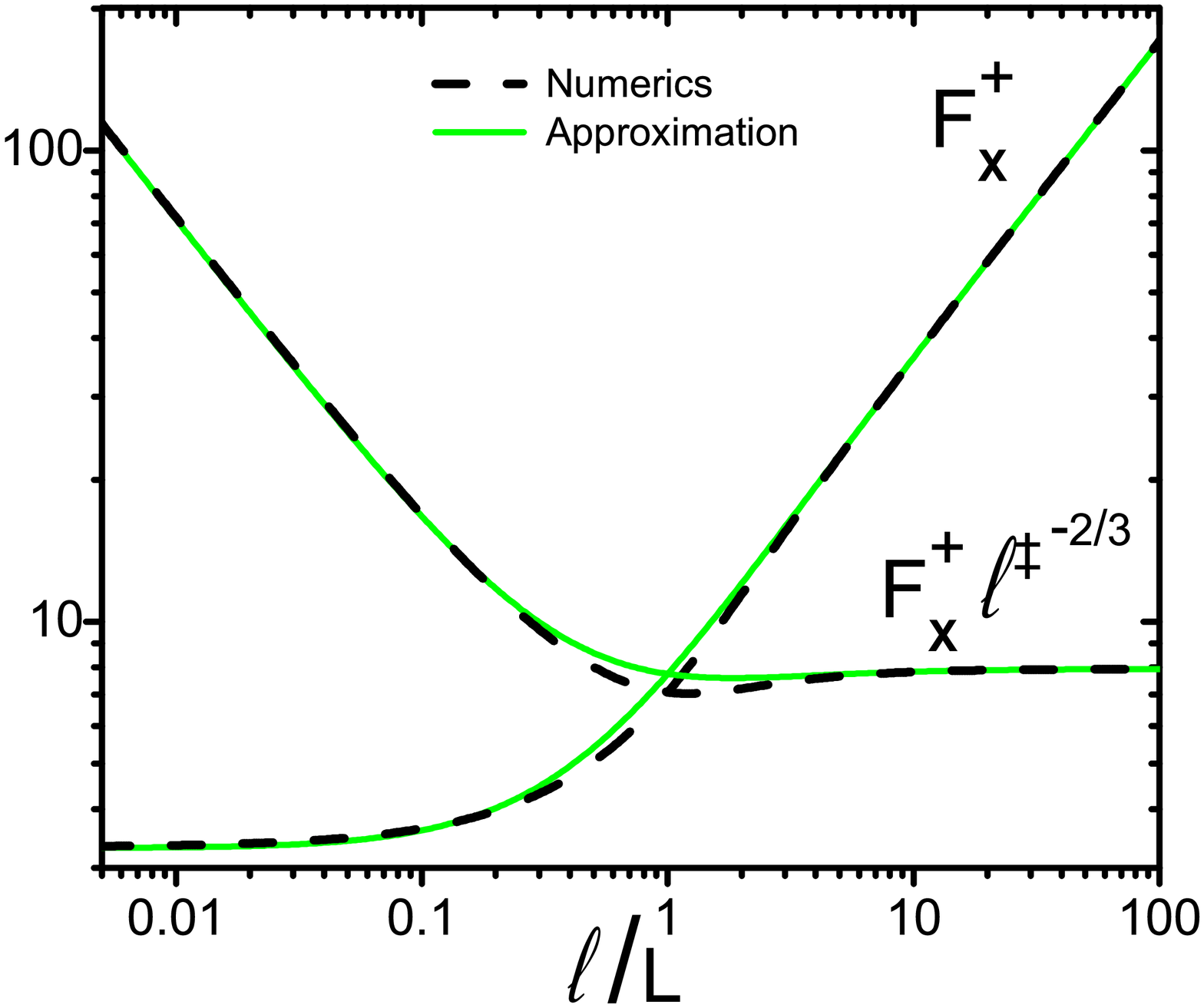}
\end{center}

  \caption{\label{f:S} Color online. Log-log plots of
  the normalized velocity mean shears $  S\Sb U
  ^\ddag \equiv  {\ell^+} \, S\Sb U^+$ and $L^+ S\Sb U^+$
  (left upper panel),  normalized mean-temperature
  gradients $ S\Sb \Theta ^\ddag \equiv   {\ell^+} \,S\Sb
  \Theta ^+ $ and  $L^+ S\Sb \Theta ^+$ (left lower
  panel), the turbulent kinetic energy  $E\Sb K^+$
  and $E\Sb K^+  /{{\ell^\ddag}}^{2/3} $ (middle upper
  panel), partial kinetic energies $\tau_{ii}/E\Sb K$
  (middle lower panel),   temperature energy $E_
  \theta ^+ $ and   $E_ \theta ^+/{{\ell^\ddag}}^{2/3}$
  (right upper panel) and horizontal thermal flux
  $F_x^+$ and   $F_x^+  /{{\ell^\ddag}}^{2/3} $ (right
  lower panel) vs. ${\ell^\ddag}=\ell/L={\ell^+}  /L^+$.
  Red and blue solid lines -- exact numerical
  solutions before and after normalization by the
  large ${\ell^\ddag}$ asymptotics, black dashed and
  dot-dashed lines -- approximate analytical
  solutions.  The region $\ell \gtrsim L$ may not be
  realized in the Nature. In this case it has only
  methodological character.    }
\end{figure*}

 This subsection is devoted to an analytical and numerical analysis of the
Minimal-Model~\eq{MM}. An example of numerical solution of Eqs.~\eq{MM}
(with some reasonable choice of the phenomenological parameters) is
shown in Fig.~\ref{f:S}. Nevertheless,  it would be much more
instructive to have approximate analytical solutions for all
correlations that will describe their ${\ell^\ddag}$-dependence with
reasonable accuracy. The detailed cumbersome procedure of finding these
solutions is skipped here, but a brief overview is as follows.

The Eqs.~\eq{MM} can be reformulated
as a polynomial equation of ninth order for the only unknown
$\sqrt{E^+\Sb{K}}$. An analysis of its structure helps to formulate an effective
interpolation formula~\eq{inter}, discussed below.
Hence, we found the solutions of Eqs.~\eq{MM} at neutral stratification, ${\ell^\ddag}=0$, corrected  up to the linear order in ${\ell^\ddag}$. Its
comparison with the existing DNS data resulted in an estimate for the
constants $\widetilde  C\Sb{RI}\approx 1.46$, and $c_{uu}\approx 0.36$.
 Then, we considered the region ${\ell^\ddag} \to \infty$. Even though such a
condition may not be realizable in nature, from a methodological
point of view, as we will see below, it enables to obtain the
desired analytical approximation. The ${\ell^\ddag} \to \infty$
asymptotic solution with corrections, linear in
the small parameter ${\ell^\ddag}^{-4/3}$, were found. Now
we are armed to suggest an interpolation formula 
\begin{subequations} 
 \label{inter} 
  {\begin{equation} \label{interA}
    {E^{+}\Sb K}({\ell^\ddag})^{3/2}\simeq \frac{11
  {\ell^\ddag}}{3 \, c_{uu}} +\frac{8\,  \widetilde C\Sb {RI}}{\sqrt{
  \big(11{\ell^\ddag}/3\, c_{uu}\big)^{2/3}
  +\big(8\,\widetilde C\Sb{RI}\big)^{1/2}}}\,, 
\end{equation} }
 that coincides
with the exact solutions for ${\ell^\ddag}=0$ and for
${\ell^\ddag}\to \infty$,  including the leading
corrections to both asymptotics, linear in ${\ell^\ddag}$,
 and ${{\ell^\ddag}}^{-4/3}$. Moreover,
in the  region  ${\ell^\ddag}\sim 1$, \Eq{interA}  accounts for the
structure of the exact polynomial. As a result, the
interpolation formula~\eq{interA} is close to the numerical solution
with deviations smaller than
 3\% in the entire region
$ 0\le {\ell^\ddag}  < \infty$, see upper middle panel on
Fig.~\ref{f:S}.  Together with \Eq{9-eqC} it produces a solution
for $S_{_U}^+$, that can be written as 
\begin{equation} 
 \label{interB}%
  S^+\Sb{U}({\ell^+} ) \simeq   \left( L_1^+ \right) ^{\!-1}+ \Big({\kappa\,
  {\ell^+}\sqrt{1+({\ell^+} /L_2^+)^{2/3}}}\,\Big)^{\!-1}\,, 
 \end{equation} 
where $ L_1^+ \equiv   3 L^+/14\,, \quad L_2^+\equiv  3L^+/{11\,\kappa} $ and
$\kappa$ is the \vK constant. This formula gives   the same accuracy
$ \sim3\%$, see upper left panel in Fig.~\ref{f:S}. We demonstrate
below that the proposed interpolation formulae describe the
${\ell^\ddag}$-dependence of the correlations with a very reasonable
accuracy, about $  10\%$,  for any $0\leq{\ell^\ddag}<\infty$,    see
black dashed lines in Figs.~\ref{f:S}.

Unfortunately, a direct substitution of the interpolation
formula~\eq{inter} into the  exact relation for $S^\ddag\Sb \Theta$
obtained from the system (\ref{MM}) works well
only for small $ {\ell^\ddag}$, in spite of the fact that the
interpolation formula is rather accurate in the whole region. We need therefore to derive an independent
interpolation formula for $S^\ddag\Sb \Theta$. Using expansions
for small $ {\ell^\ddag}  \ll 1$ and large $
{\ell^\ddag}  \gg 1$ we suggest   
{ \begin{eqnarray}\label{inter1a}
  S\Sb \Theta ^+ ({\ell^+})  \simeq  {S\Sb \Theta ^+}^\infty
\!\! + \frac{S^+\Sb {\Theta,{\scriptstyle 0}}\!+\! 6
(c_{uu}\alpha)^{4/3}S\Sb {\Theta,1}^{+\infty}} { \left( 1 +\alpha\,{\ell^+}/L^+
 \right) ^{4/3}}\,,~~~~~~
\end{eqnarray}}
in which
\begin{eqnarray}\nonumber
S^+ \Sb {\Theta,\scriptstyle 0}  &=& 2^{1/4}c_{uu} C\Sb{U\Theta}/
\widetilde C\Sb{RI}^{1/4 }\ell ^+,\\ \nonumber
S^{+\infty}\Sb {\Theta,1} &=& -{C_{u\theta}}(2\widetilde C\Sb{RI}
-{(11\,C_{u\theta} -3\,C\Sb{SU})}/{{3\,S\Sb \Theta ^{+\infty}}L^+}
)/{L^+},\\ \nonumber
S\Sb \Theta ^{ + \infty} &=& -14 (C\Sb{SU}- 4 C\Sb{U\Theta}/3)/3
L^+,
 \end{eqnarray}
 and $\alpha$ satisfies
 \begin{eqnarray}\nonumber
S^+\Sb {\Theta,{\scriptstyle 1}}{\ell^+} &=& {S\Sb \Theta ^+}^\infty
L^+ + 6S\Sb {\Theta,1}^{+\infty}L^+ (c_{uu}\alpha)^{4/3} -{4 \alpha S\Sb
{\Theta,{\scriptstyle 0}}}/3 \,,
\end{eqnarray}
with
 \begin{eqnarray}\nonumber
 S^+\Sb {\Theta,
{\scriptstyle{1}}} {\ell^+} &=& -{C_{u\theta}}  \left( 3/{4\widetilde C\Sb{RI}} -22
+3 {C\Sb{SU}}/{C_{u\theta}} \right) / {24\widetilde C\Sb{RI}}\ .
\end{eqnarray}
\end{subequations}
 Equation~\eq{inter1a} is constructed such that the leading and
 sub-leading asymptotics for small and large $ {\ell^\ddag} $ coincide
 with the first two terms in the exact expansions at "almost" neural stratification
 and extremely strong stratification. As a result, \Eq{inter1a} approximates the exact solution
 with errors smaller then 5\% for $ {\ell^\ddag} < 1$ and $ {\ell^\ddag} > 50$
 and with errors smaller than 10\% for any $ {\ell^\ddag} $, see
 lower left panel in \Fig{f:S}.

 Substituting the approximate Eqs. \eq{inter}   into
 the exact relations  \eq{MM}, one gets
 approximate solutions $E^+_\theta$ and
 $F_x^+$ with errors smaller than 10\%, see rightmost panels
 in \Fig{f:S}.

\subsection{\label{ss:one} Mean velocity and temperature profiles}

In principle, integrating the mean shear $S\Sb U^+$ and the mean
temperature gradient $S\Sb \Theta ^+$, one can find the mean
velocity and temperature profiles. Unfortunately, to do so we need
to know $S\Sb U^+ $ and $S\Sb \Theta ^+$ as functions of the
elevation $z$, while in our approach they are found as functions of
$\ell/L$. Remember, that the external parameter $\ell$ is the outer
scale of turbulence that depends on the elevation $z$.
The importance of an accounting for the proper physically motivated dependance of $\ell$ on $z$ for an example of channels and pipes has been recently shown by L'vov et al. (2008).
For the problem at hands, we can safely take $\ell=z$ if $z\ll L$, however
when $z>L$  the function $\ell(z)$ is not found theoretically
although it was discussed phenomenologically with support of
observational, experimental and numerical data. It is traditionally
believed that for $z \gtrsim L$ the scale $\ell$  saturates at some
level of order $L$ [see, e.g. Eq. \Ref{sless}].

\begin{figure*}
 \includegraphics[width=0.475  \textwidth]{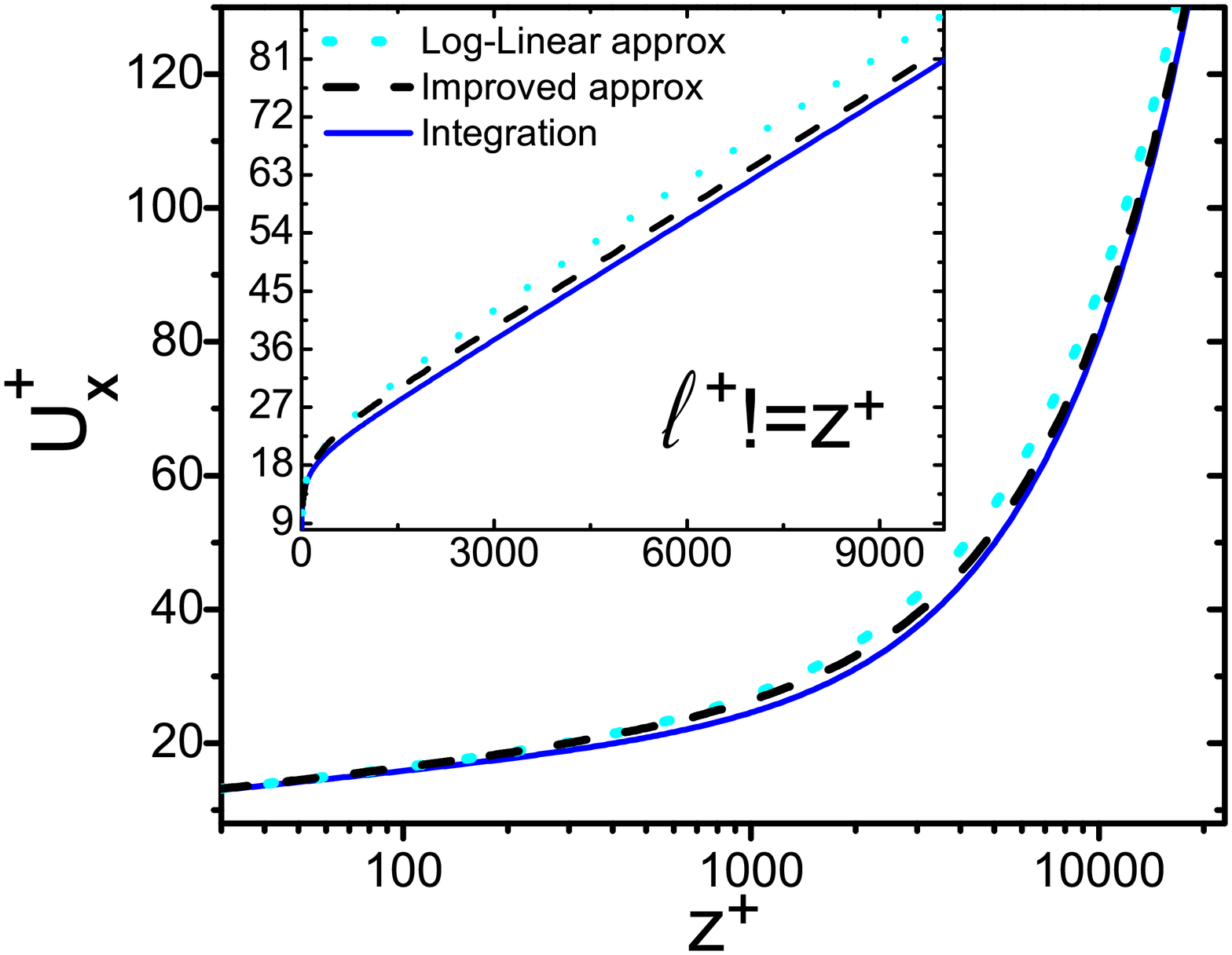}
 \includegraphics[width=0.475  \textwidth]{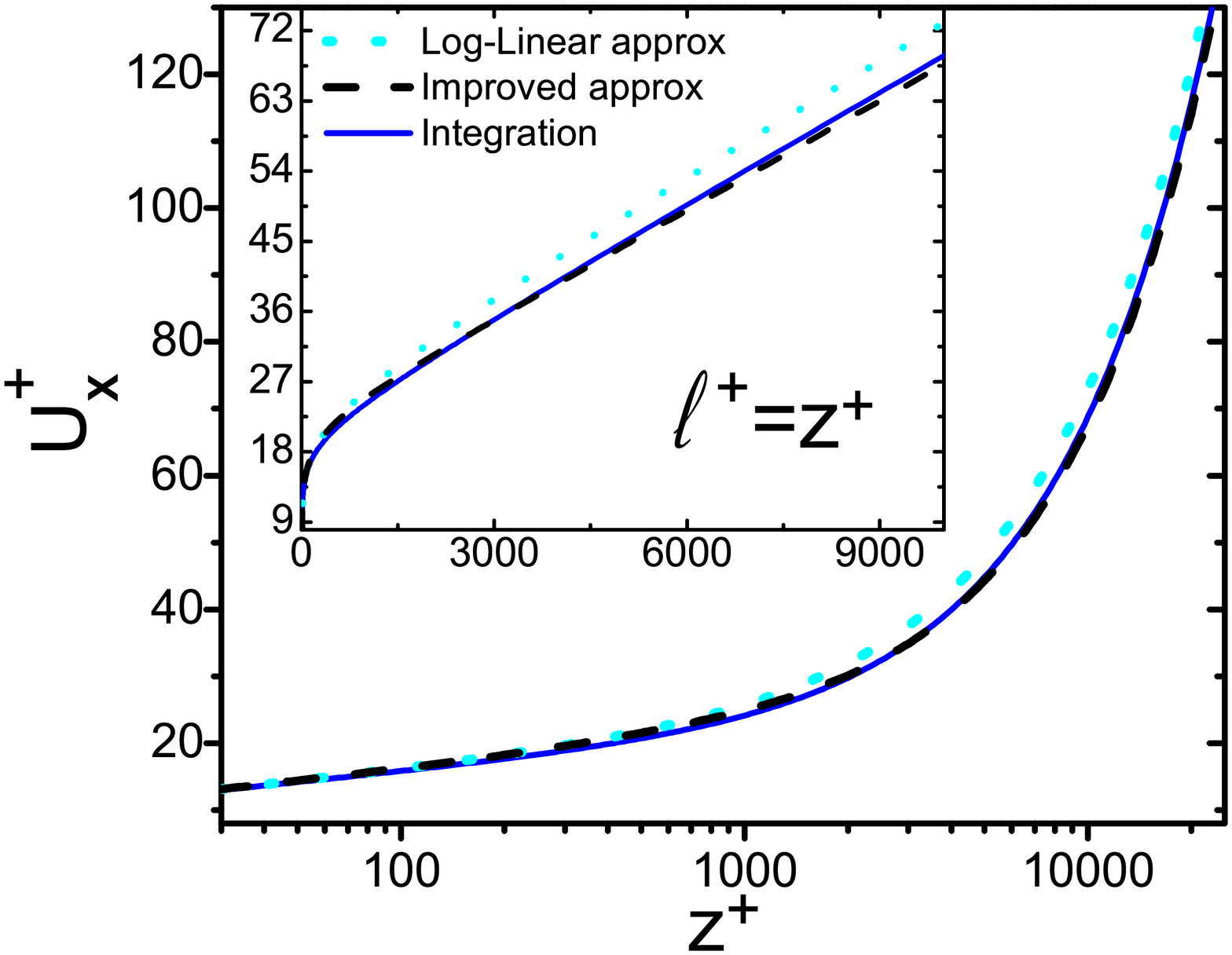}
 \caption{\label{f:means}
 Computed with \Eq{sless} (for $d_1=d_2=1$) plots of $ U^+ $
 (blue solid lines) vs $\ln(z/L)$  and vs. $ z/L $ (inserts) for $L^+=1000$. In
 the left panel $\ell(z)$ is taken from \Eq{sless}, while in the right
 panel $\ell(z)=z$.   Log-linear approximation~\eq{ll-aprV}
 to all profiles is shown by dotted lines, its improved
 version~\eq{ll-aprV2} by dashed lines. ``$!\!=$" stands for $\neq$. The region $\ell \gtrsim L$ may not
  be realized in the Nature. In this case it has only
  methodological character.  }
\end{figure*}

The resulting plots of $U^+$  are  shown on \Fig{f:means}, left
panel. Even taking $\ell(z)=z$ one gets a very similar velocity
profile, see \Fig{f:means}, right panel. With $\ell(z)=z$ we found an
analytical expression for the mean-velocity profile using the
interpolation \Eq{interB} for $S\Sb U^+$: 
\begin{equation}  
\label{vel} 
  U^+(z)=\frac{1}{\kappa}\ln\!\Bigg[\frac{z/z_{u0}}{\Big( 1+\sqrt{1+  \left( z /L _2 \right)^{2/3}}\ \Big)^{\!3}} \Bigg]+\frac{z }{L_1 }  \ . 
\end{equation}  
Here $z_{u0}$ is the roughness length.

The resulting mean velocity profiles have  logarithmic asymptotic
for $z<L$ and a linear behavior for $z>L$ in agreement with
meteorological observations. Usually the observations are
parameterized by a so-called log-linear approximation (Monin and Obukhov, 1954):
\begin{subequations} 
\label{ll-apr} 
\begin{equation}  
  \label{ll-aprV}  
  U^+ =  {\kappa}^{-1}  \ln ({z}/{z_{u0}})+ {z}/{L_1}\,,
\end{equation}
which is plotted in \Fig{f:means} by dotted lines. One sees some
deviation in the region of intermediate $z$. The reason is that the
real profile [see, e.g. \Eq{vel}] has a logarithmic term that
saturates for $z\gg L$, while in the approximation~\eq{ll-aprV} this
term continues to grow. To fix this one can use \Eq{vel} (with
$L_2=L_1$ for simplicity), or even its simplified version \begin{equation}
\label{ll-aprV2}  
U^+ = \frac{1}{\kappa}  \ln \frac{z}{z_{u0}\sqrt{1+(z/L_1)^2}}+
 \frac{z}{L_1}\ .
 \end{equation} \end{subequations}
This approximation is plotted as
a dashed line on \Fig{f:means} for comparison.
 One sees that the  approximation~\eq{ll-aprV2} works
much better than the traditional one. Thus we suggest \Eq{ll-aprV2}
for parameterizing meteorological observations.

The temperature profiles in our approach look  similar to the velocity
ones: they have logarithmic
asymptotic for $\ell < L$ and linear behavior for $\ell >L$.
Correspondingly, they can be fitted by a log-linear
approximation, like~\eq{ll-aprV}, or even better, by an improved version of
it, like \Eq{ll-aprV2}. Clearly, the values of constants will be
different:  $\kappa \Rightarrow \kappa\Sb T$, $L_1 \Rightarrow
L_{1,\rm T}$, etc.
\subsection{\label{ss:one}  Profiles of second-order correlations}

The computed profiles of the turbulent kinetic and
 temperature energies, horizontal thermal flux profile and the
 anisotropy profiles are shown on \Fig{f:S} in the middle and right
 panels. The anisotropy profiles, lower middle panel, saturate at
 $\ell/L\approx 2$, therefore they are not sensitive to the $z$-dependence of $\ell(z)$;
 even quantitatively one can think of these profiles as if they were plotted as a function of $z/L$.

 Another issue is the profiles of $E\Sb K^+$ (upper middle panel) and
 of $E\Sb \Theta$ and $F_x^+$ (rightmost panels), that are $\propto (\ell/L)^{2/3}$ for
 $\ell\gg L$ (if realizable). With the interpolation formula~\eq{sless} the
 profiles of the second order correlations have to saturate at
 levels corresponding to ${\ell^\ddag} = 1$. This sensitivity to the $z$-dependence of
 $\ell(z)$ makes a comparison of the prediction with
 experimental data very desirable.

 \subsection{\label{ss:nums} Turbulent transport, Richardson and Prandtl numbers}

In our notations the turbulent viscosity and thermal conductivity,
turbulent Prandtl number, the gradient- and flux-Richardson numbers are 
\begin{subequations} \label{num}
  \begin{eqnarray} \label{nu} 
  \nu \Sb T&\equiv  & - \frac{\tau_{xz}}{S\Sb U}= \frac1 {S\Sb U ^+}\equiv  C_\nu ({\ell^\ddag}  )
  \frac{\tau_{zz}^+}{\gamma  _{uu}^+}\,,\\
   \label{chi}
    \chi\Sb T &\equiv & -\frac{F_z}{S\Sb \Theta}= \frac 1 {S\Sb \Theta ^+}
    \equiv  C_\chi  ({\ell^\ddag}  )
  \frac{\tau_{zz}^+}{\gamma  _{uu}^+}\,,
     \\  \label{numPr}
\mbox{Pr}\Sb T& \equiv  &\frac{\nu\Sb T }{\chi\Sb T }
=\frac{S_{_\Theta}^+}{S_{_U}^+}
=\frac{S_{_\Theta}^\ddag}{S_{_U}^\ddag}\,,  \\
\Rig  &\equiv & \frac{\beta  S_{_\Theta}}{S_{_U}^2}=  \frac{
S_{_\Theta}^+}{L^+\, {S_{_U}^+}^2 }=\frac{{\ell^\ddag}
S_{_\Theta}^\ddag}{{S_{_U}^\ddag}^ 2}\,, \\
\Rif  &\equiv & \frac{\beta   F_z}{\tau_{xy}S_{_U}}=  \frac1{ L^+\, S_{_U}^+
}=\frac{{\ell^\ddag}   }{ S_{_U}^\ddag }\,, \label{numRig}\\
\label{rel4} \Rig&=&\Rif \, \mbox{Pr}\Sb T\ .
\end{eqnarray}
 \end{subequations} 
With \Eqs{nu} and \eq{chi} we introduce also two dimensionless
functions $C_\nu ({\ell^\ddag}  )$  and $C_\chi  ({\ell^\ddag}  )$ that are taken as
${\ell^\ddag} $-independent constants in the down-gradient transport
approximation~\eq{dt} described in the Introduction.  We will show, however,  that these functions
have a strong dependence on ${\ell^\ddag} $, going to zero in the limit
${\ell^\ddag} \to\infty$ as $1/ {\ell^\ddag} ^{4/3}$. Therefore this approximation is
not valid for large ${\ell^\ddag} $ even qualitatively.

\subsubsection{\label{sss:dt} Approximation of down-gradient transport
and its violation in stably stratified TBL}
\begin{figure*}
\begin{center}
 \includegraphics[width=0.295 \textwidth]{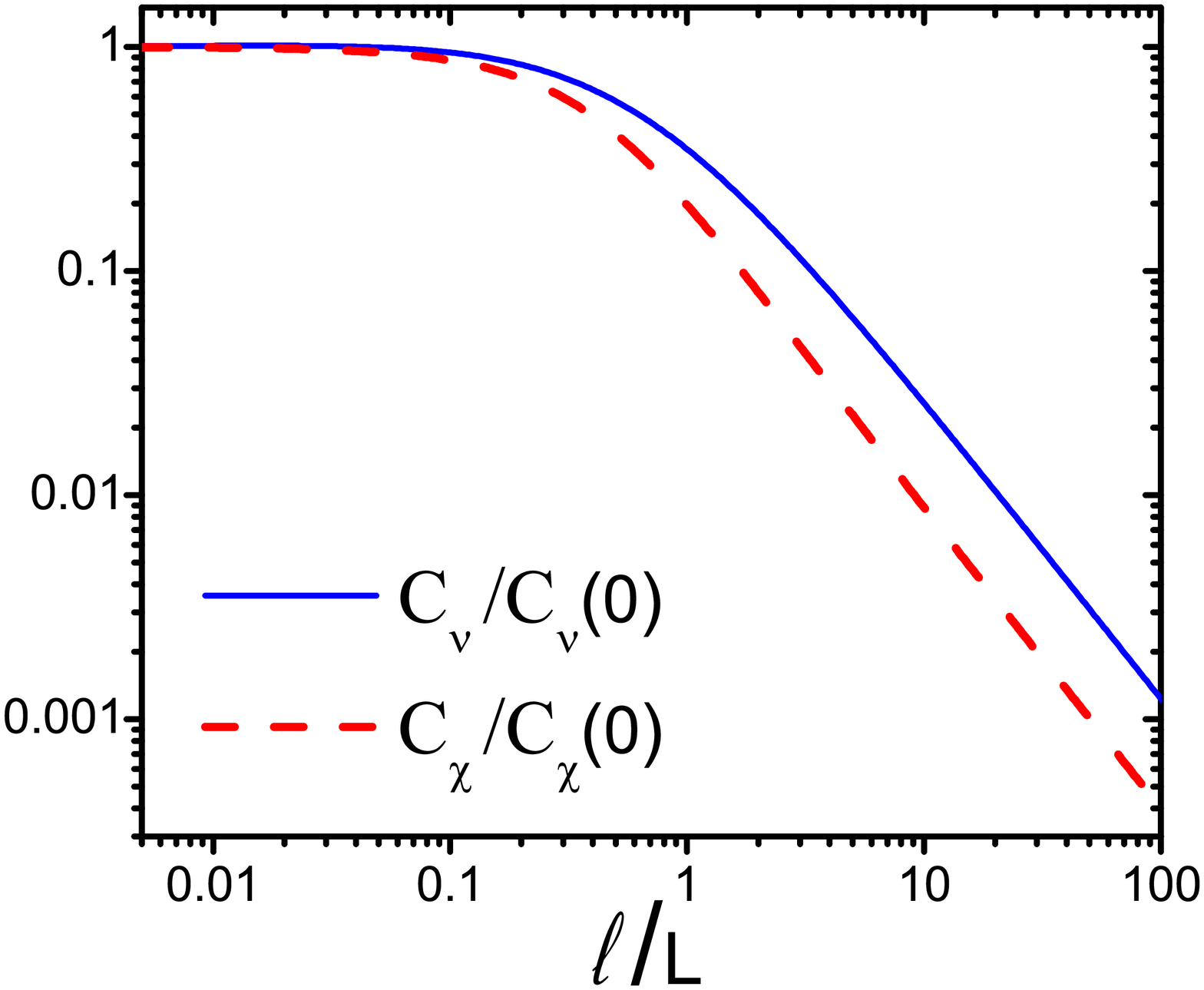}
 \includegraphics[width=0.317  \textwidth]{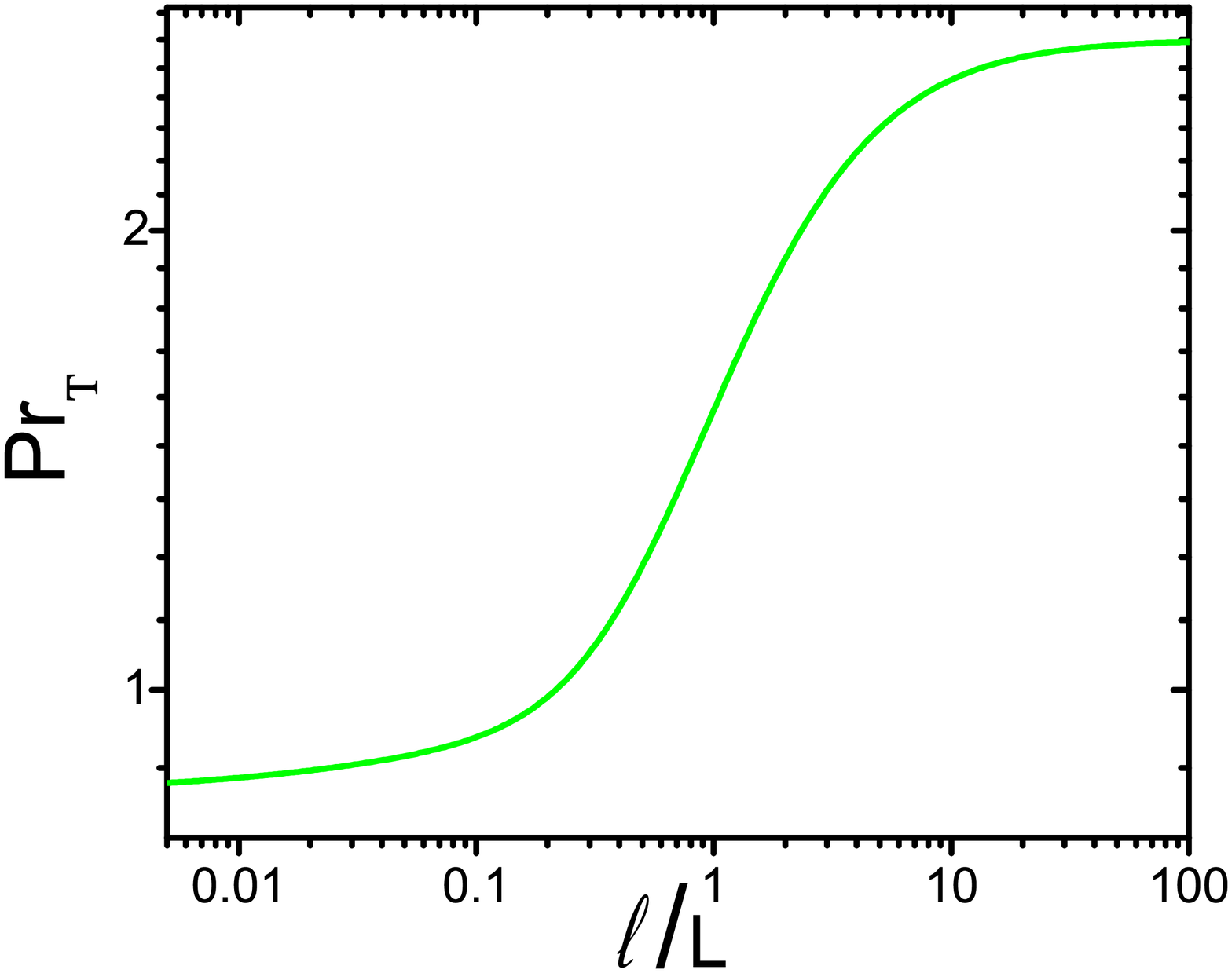}
 \includegraphics[width=0.3  \textwidth]{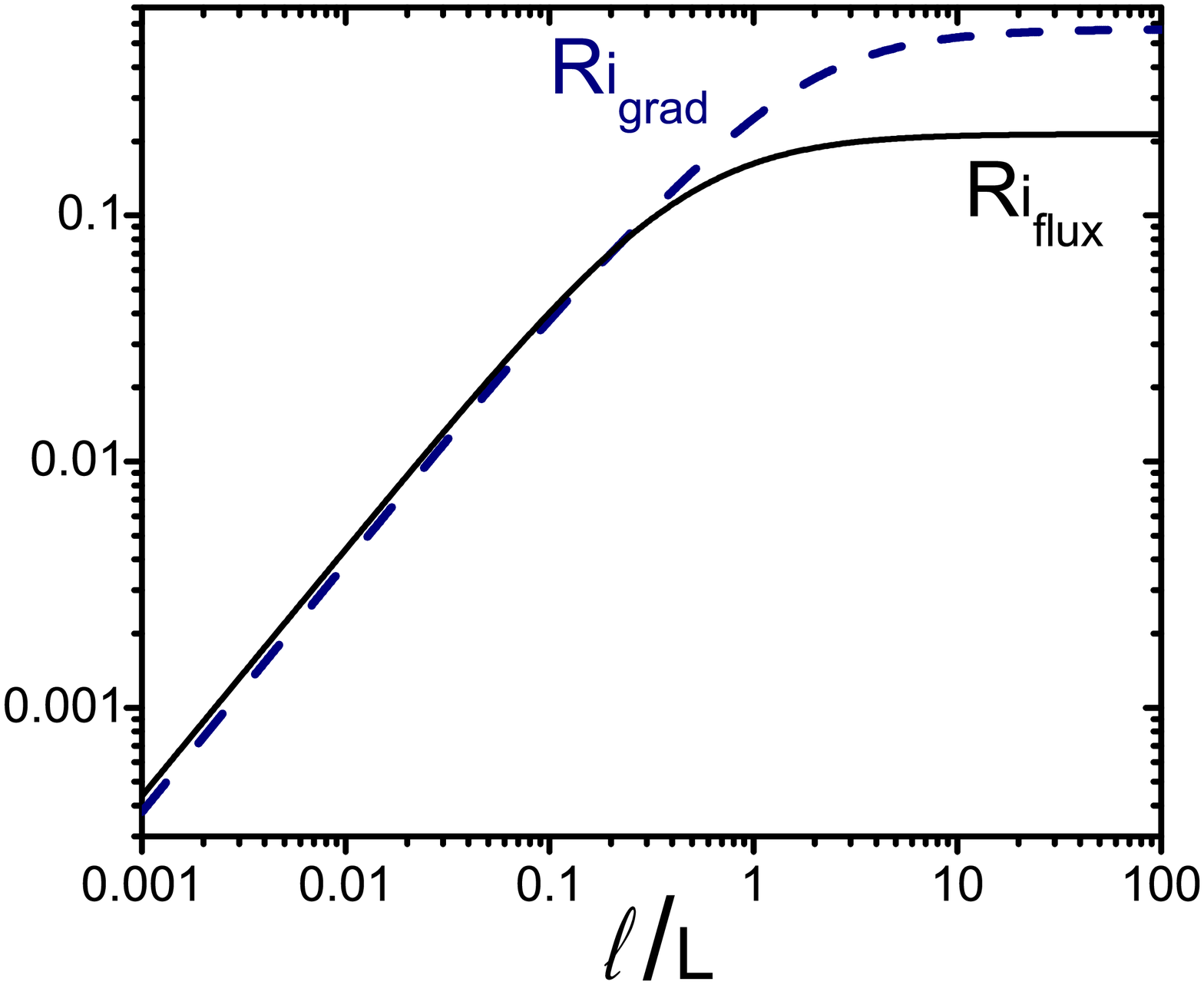}\\
 \includegraphics[width=0.305  \textwidth]{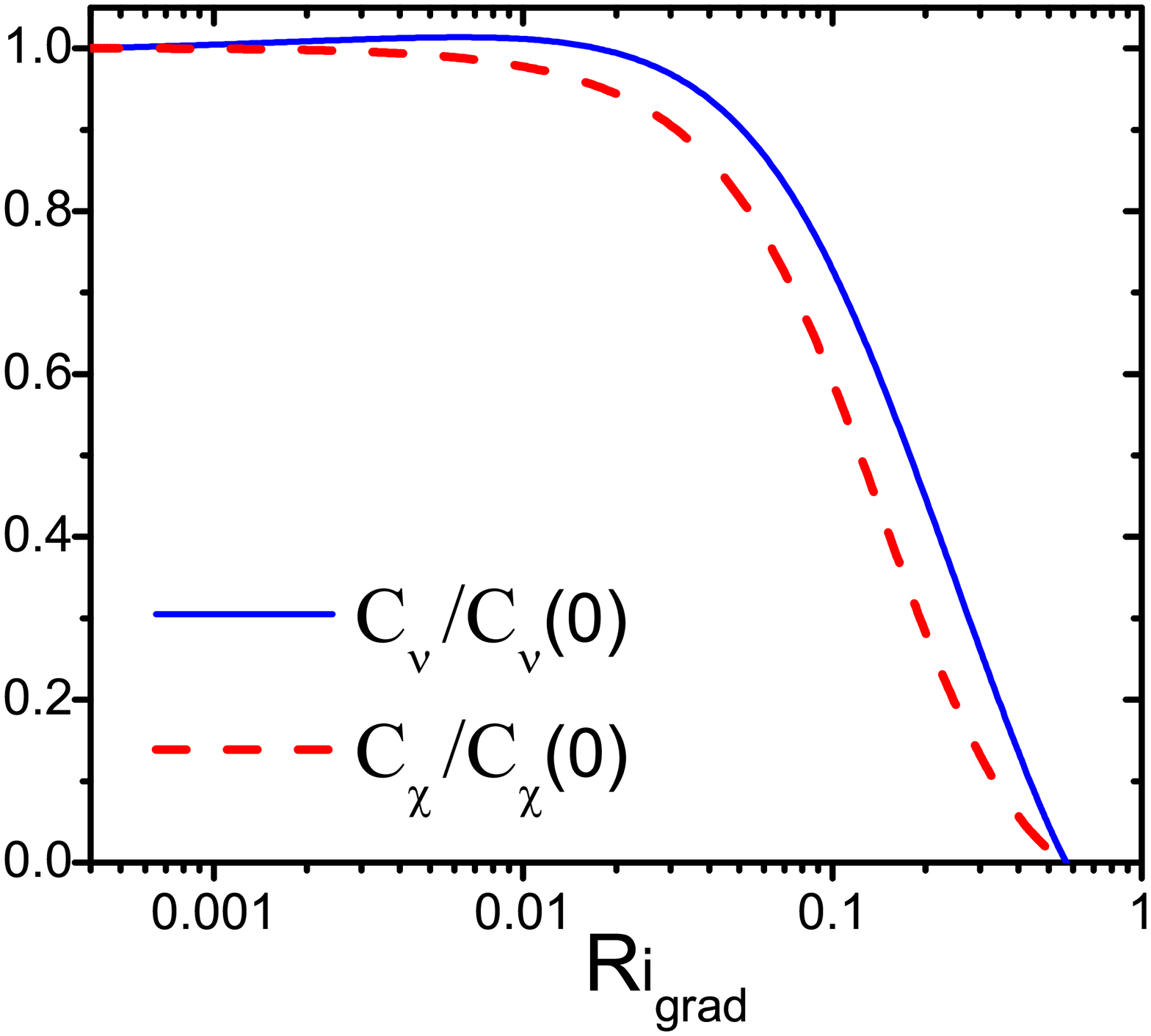}
 \includegraphics[width=0.312  \textwidth]{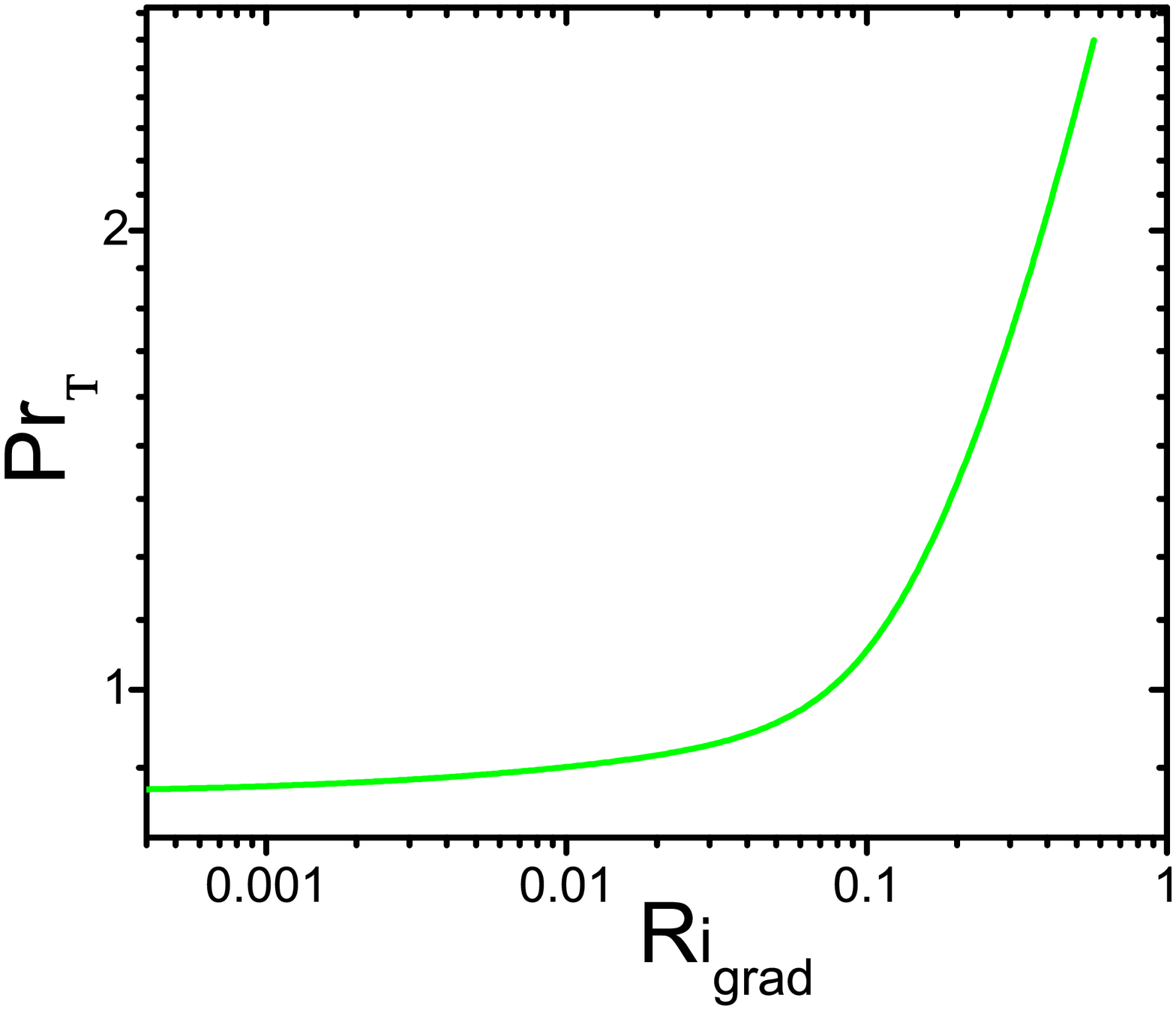}
 \includegraphics[width=0.301  \textwidth]{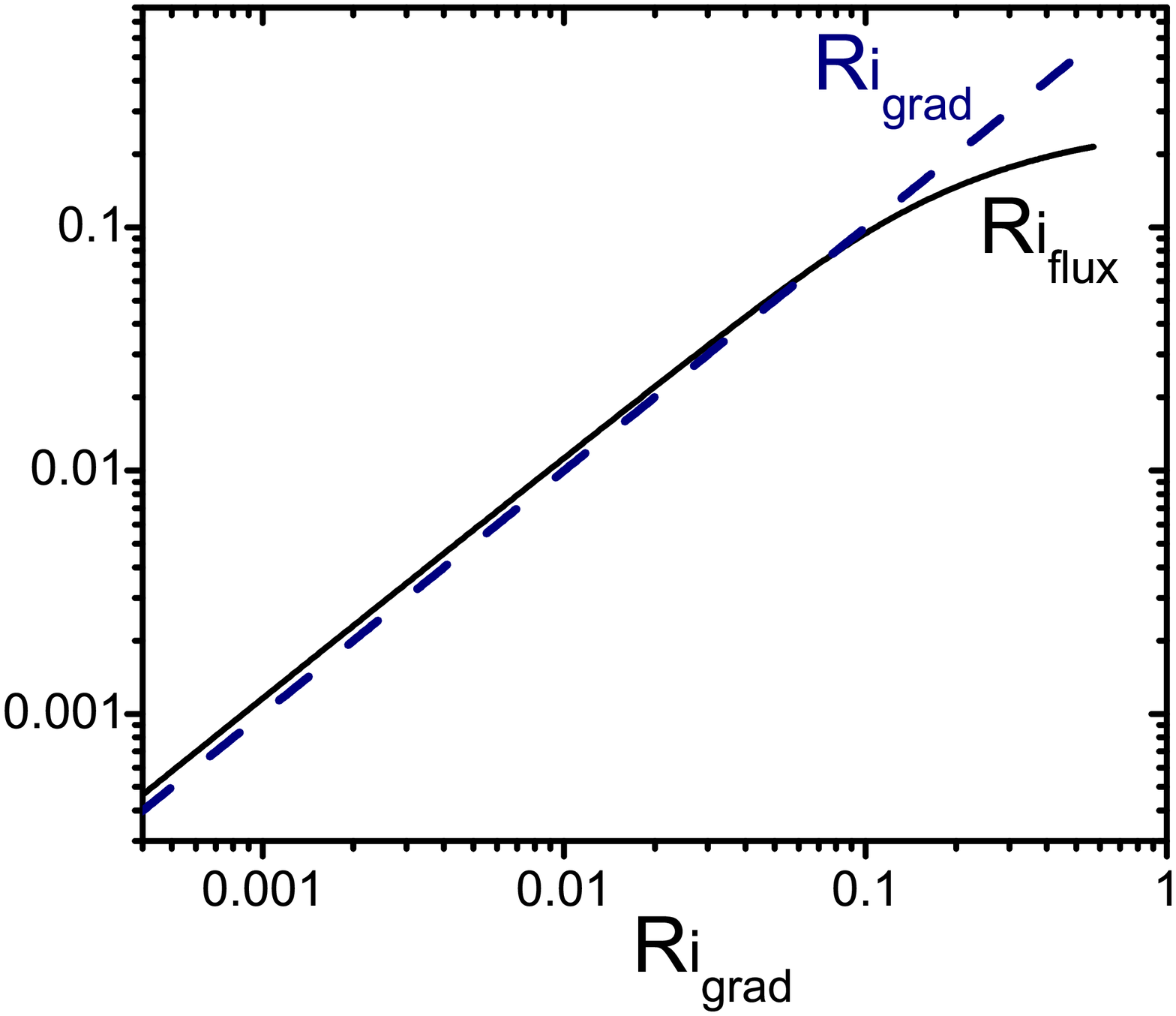}
\end{center}

 \caption{\label{f:res}  Color online.
 Log-log plots of ``down-gradient coefficient-functions"
    $C_\nu$ (solid blue lines) and
   $C_\chi$ (red dashed lines) --  left panels;
    turbulent Prandtl number Pr$\Sb T$  (green lines on middle panels)
   and $\Rif$ (black solid lines),  $\Rig$ (black dashed lines) -- on right panels
   as function of  ${\ell^\ddag} =\ell/L$ (upper panels)
   and vs. $\Rig$ (lower panels). Notice, that the presented dependencies
   have qualitative character, and the choice of constants $C_{...}$
   depends on the actual functional form $\ell \left( z \right) $. For simplicity,
   we took $\ell \left( z \right)  = z$.The region $\ell \gtrsim L$ may not be
  realized in the Nature. In this case it has only
  methodological character.  }
\end{figure*}
 As we mentioned in the Introduction, the concept of the
 down-gradient transport assumes that the momentum and thermal
 fluxes are proportional to the mean velocity and temperature
 gradients, see \Eqs{dt}:
  \begin{equation} \label{appr}
 \tau_{xz}=-\nu\Sb T  S_{_U}\,,\quad
 F_z=-\chi\Sb T  S\Sb \Theta \,,
 \end{equation}
 where $\nu\Sb T$ and $\chi\Sb T$ are effective turbulent viscosity
 and thermal conductivity, that can be estimated  by dimensional
 reasoning. Equations \eq{dt}, giving this estimates,  include
 additional physical arguments that vertical transport parameters
 should be estimated via vertical turbulent velocity,
 $\sqrt{\tau_{zz}}$, and characteristic vertical scale of turbulence,
 $\ell_z$. The relations between the scales $\ell_j$ in different
 $j$-directions in anisotropic turbulence can be found in the
 approximation of time-isotropy, according to which
 \begin{equation} \label{ti}
 \frac{\sqrt{\tau_{xx}}}{\ell_x}=\frac{
 \sqrt{\tau_{yy}}}{\ell_y}= \frac{\sqrt{\tau_{zz}}}{\ell_z}\equiv
 \gamma\  \Rightarrow \ \gamma_{uu}\ .
\end{equation} 
 Here $\gamma$ is a characteristic isotropic frequency of turbulence,
 that for concreteness can be taken as the kinetic energy relaxation
 frequency $\gamma_{uu}$.
The approximation~\eq{ti} is supported by experimental data, according
to which in anisotropic turbulence the ratios $\ell_i/\ell_j$ ($i\ne
j$) are larger then the ratios $\ell_i\, \sqrt{\tau_{jj}}\,
/\ell_j\, \sqrt{\tau_{ii}}$ that are close to unity. With this
approximations $\nu\Sb T$ and $\chi\Sb T$ can be estimated as
follows:
\begin{equation} \label{est1} \nu\Sb T= C_\nu  {\tau_{zz}}/{\gamma_{uu}}\,, \quad
\chi \Sb T = C_\chi {\tau_{zz}}/{\gamma_{uu}}\,, \end{equation}  
where, according to the approximation of down-gradient transport,
the dimensionless parameters $C_\nu$ and $C_\chi$ are taken as
constants, independent of the level of stratification.

In order to check how the approximation~\eq{appr},~\eq{est1} works
in the stratified TBL  for both fluxes, one can consider \Eqs{appr}
as \emph{definitions} of  $\nu\Sb T$ and $\chi\Sb T$ and
\Eqs{est1} as \emph{definitions} of  $C_\nu$ and   $C_\nu$.
This gives
\begin{subequations} \label{defC} \begin{eqnarray} \label{defCM} 
C_\nu&\equiv & -\frac{\tau_{xz}}{\tau_{zz}}\, \frac{\gamma_{uu}}{S_{_U}}=
\frac{\gamma_{uu}^+}{ \tau_{zz}^+ S_{_U}^+} \,, \\  \label{defCH}
C_\chi&\equiv & -\frac{F_x}{\tau_{zz}}\, \frac{\gamma_{uu}}{S\Sb \Theta}=
\frac{\gamma_{uu}^+}{ \tau_{zz}^+ S\Sb \Theta^+} \ .  \end{eqnarray} \end{subequations}
Recall, that in this paper the down-gradient approximation is not
used at all. Instead, we are using exact balance equations for all
relevant second order correlations, including $\tau_{xz}$ and $F_x$.
Substituting our results in the RHS of the definitions~\eq{defC} we
can find, how $C_\nu$ and $C_\chi$ depend on ${\ell^\ddag} =\ell/L$ that
determines the level of stratification in our approach.

The resulting plots of the ratios  $C_\nu({\ell^\ddag} )/C_\nu(0) $ and
$C_\chi({\ell^\ddag} )/C_\chi(0)$ are shown in the leftmost panel in
\Fig{f:res}. One sees that the $C_\nu({\ell^\ddag} )$ and $C_\chi({\ell^\ddag} )$ can be
considered approximately as constants only for $\ell \le 0.2\, L$.
For larger $\ell/ L$ both $C_\nu({\ell^\ddag} )$ and $C_\chi({\ell^\ddag} )$ rapidly
decrease, more or less in the same manner, diminishing by an order of magnitude
already for $\ell \approx 2\, L$. For larger $\ell/L$ one can use the
asymptotic solution according to
which 
\begin{equation}  \label{est2}
 S_{_U}^+\simeq \frac1{L^+}\,,\  \gamma_{uu}\simeq
\frac{\sqrt{E\Sb K^+}}{{\ell^+} }\simeq \frac{{\ell^\ddag} ^{1/3}}{{\ell^+} }\,, \
\tau_{zz}\simeq {\ell^\ddag} ^{2/3}\ . 
\end{equation} 
This means that  both functions vanish as $1/{\ell^\ddag} ^{4/3}$: 
 \begin{equation} \label{as}
C_\nu({\ell^\ddag}  )\simeq 0.01  \left(  \frac{L}{ \ell}  \right) ^{4/3}\!\!  ,  \ \
C_\chi({\ell^\ddag} )\simeq 0.003    \left(  \frac{L} { \ell}  \right) ^{4/3} \!\!  ,
 \end{equation}  
 where numerical prefactors account for the accepted values of
 the dimensionless fit parameters.

The physical reason for the strong dependence  of $ C_\nu$ and
$C_\chi$ on stratification is as follows: in the   RHS of \Eq{txz}
for the momentum flux and  \Eq{Sim-F-z} for the vertical heat flux
there are two terms. The first ones, proportional to $\tau_{zz}$ and
velocity (or temperature) gradients
  correspond to the   approximation~\eq{appr}, giving
(in our notations) $C_\nu =$const and $C_\chi =$const, in agreement
with the down-gradient transport concept. However, there are second
contributions to the vertical momentum flux $\propto F_x$ and to the
vertical heat flux,  that is proportional to $ \beta  E_\theta$. In our
approach both contributions are negative, giving rise to the
\emph{counter-gradient fluxes}.   What follows from our   approach,
is that these counter-gradient fluxes cancel (to the leading order) the
down-gradient contributions in the limit ${\ell^\ddag} \to\infty$. As a
result, in this limit the effective turbulent diffusion  and thermal
conductivity   vanish, making the down-gradient approximation for
them (with constant $C_\nu$ and $C_\chi$) irrelevant even
qualitatively for   $\ell \gtrsim L$.

 In our picture
of stable temperature-stratified TBL, the turbulence exists at any
elevations, where one can neglect the Coriolis force. Moreover, the
turbulent kinetic and temperature energies increase as
$(\ell/L)^{2/3}$ for $\ell > L$, see
\Fig{f:S}.   At the same time, the mean velocity and potential
temperature change the $(\ell/L)$-dependence from logarithmic lo
linear, see \Fig{f:means} and (modified) log-linear interpolation
formula~\eq{ll-aprV2}. Correspondingly, the shear of the mean
velocity and \emph{the mean temperature gradient} saturate at some
elevation (and at some $\ell/L$), and $\Rig$\emph{ saturates as
well}. This predictions agree with large eddy simulation by
Zilitinkevich and Esau (2006), where
$\Rig$ can be considered as saturating around 0.4 for $z/L\approx
100$.

 Notice that the turbulent closures of kind used above
cannot be applied for strongly stratified flows with $\Rig\gtrsim 1$
(may be even at  $\Rig\sim 1$). There are two reasons for that.  The
first one was mentioned in the Introduction. Namely, for
$\Rig\gtrsim 1$ the Brunt-V\"ais\"al\"a frequency $ N\equiv \sqrt{\beta  S\Sb
\Theta}$, $ N^+= \sqrt{{S\Sb \Theta^+}/{L^+}}$,
 is   larger then the eddy-turnover frequency  and
 therefore   there are weakly decaying Kelvin-Helmoholtz internal
 gravity
 waves which, generally speaking, have to be accounted for in the momentum
 and energy balance equations.

 The second reason, that makes the results very sensitive to the
 contribution of internal waves follows from the fact that
 vortical  turbulent fluxes vanish (at fixed velocity and temperature
 gradients). Therefore even relatively small contributions of
 different nature to the momentum and thermal fluxes may be important.

 The final conclusion is that the TBL modeling at large level of
 stratification requires an accounting for turbulence of the internal
 waves together with the vortical turbulence. Definitely, new
 observations, laboratory and numerical experiments with control of
 internal wave activity are very likely.


\acknowledgements 
VL kindly acknowledge   the possibility to give an invited lecture on the problems, discussed in this paper,  at the International Conference "Turbulent Mixing and Beyond", which was held August 2007 at the Abdus Salam International Center for Theoretical Physics, Trieste, Italy.
 VL  and OR also acknowledge the support of the  Transnational Access Programme at RISC-Linz, funded by the European Commission Framework 6 Programme for Integrated Infrastructures Initiatives under the project SCIEnce (Contract No. 026133).

\appendix

\section{\label{ss:onClosure} On the closure problem of triple  correlations via second
order correlations}

Let us look more carefully at the approximation~\eq{freqs}, which is
\begin{eqnarray}\label{App-Appr}
\gamma _{uu}&=&c_{uu} \sqrt{E\Sb K} \big / \ell\,, \quad \gamma \Sb {RI}= C\Sb
{RI} \gamma _{uu}\,,  \\ \nonumber \widetilde  \gamma\Sb {RI}& = &  \widetilde  C\Sb {RI} \gamma\Sb
{RI}\,, \gamma_{\theta\theta}  = C_{\theta\theta} \gamma _{uu}\,, \quad
\gamma\Sb{RD}= C_{u\theta}\gamma_{uu}\ .
\end{eqnarray}
The dimensional reasoning that
leads to this approximation is questionable for problems having a
dimensionless parameter ${\ell^\ddag} $.  Generally speaking, all
``constants" $c_{...}$ and $C_{...}$ in \Eq{App-Appr} can be any
functions of ${\ell^\ddag} $.  Presently we just  hope that a possible ${\ell^\ddag} $
dependence of these functions is relatively weak and does not
affect the qualitative picture of the phenomenon.

Moreover, even the assumption  \eq{distau1a} that the dissipation
of the thermal flux $\epsilon _i$ is proportional to the thermal flux and
the assumption \eq{distau1b} that the dissipation of $E_\theta$,
$\varepsilon \propto E_\theta$ are also questionable. Formally speaking,
one cannot guarantee that the triple cross-correlator $  \left\langle \theta
uu  \right\rangle ^+$ that estimates $\epsilon ^+$, can be (roughly speaking)
decomposed like $ \left\langle u\theta \right\rangle  \sqrt { \left\langle uu \right\rangle }$, i.e  really
proportional to $F= \left\langle u\theta \right\rangle $ as it stated in   \Eq{distau1a}.
Theoretically, one cannot exclude the decomposition $ \left\langle \theta uu
 \right\rangle \sim  \left\langle uu \right\rangle  \sqrt { \left\langle \theta\theta \right\rangle }$, i.e. a contribution to
$\epsilon \propto E \Sb K$. Similarly, the dissipation $\varepsilon $ in the
balance~\eq{corrc} of $E_\theta$, that is determined by the
correlator~\eq{distau1b}, is $\propto  \left\langle \theta\theta u  \right\rangle $, as it
follows from the decomposition $ \left\langle \theta\theta u  \right\rangle \sim
 \left\langle \theta\theta \right\rangle  \sqrt { \left\langle uu \right\rangle } $ and is stated in \Eq{distau1b}.
This correlator allows, for example, the decomposition
$ \left\langle \theta\theta u  \right\rangle \sim  \left\langle \theta u \right\rangle  \sqrt { \left\langle \theta\theta \right\rangle } $,
i.e. contribution to $\varepsilon \propto F$. This discussion demonstrates,
that the situation with the dissipation rates is not so simple, as
one may think and thus requires careful theoretical analysis that
is in our agenda for future work. Our preliminary analysis of this
problem shows that all fitting constants are indeed functions of
${\ell^\ddag} $. Fortunately, they vary within finite limits in the entire
interval $0\le {\ell^\ddag}  < \infty$. Therefore we propose that the
approximations used in this paper preserve the qualitative picture
of the phenomenon. Once again, the traditional down-gradient
approximation does not work even qualitatively because
corresponding ``constants" $C_\nu$ and $C_\chi$ vanish in the
limit ${\ell^\ddag} \to\infty$.


%


\begin{thebibliography}{99}

\bibitem{1903Bou} Boussinesq, J.: 1903, The'orique Analytique
de la Chaleur, Vol. 2. Gauthier-Villars, Paris.

\bibitem{Cheng2002} Cheng, Y., Canuto, V. M., and Howard, A. M., 2002: An improved model for the turbulent PBL, \emph{J. Atm. Sci.}, \textbf{59}, 1550-1565.

\bibitem{Canuto2002} Canuto, V. M., 2002: Critical Richardson numbers and gravity waves, \emph{Astronomy} \& \emph{Astrophysics}, \textbf{384}, 1119-1123.

\bibitem{Elperin02} Elperin, T., Kleeorin, N., Rogachevskii, I., and Zilitinkevich, S., 2002: Formation of large-scale semi-organized structures in turbulent convection. \emph{Phys. Rev. E}, \textbf{66}, 066305.

\bibitem{Galperin2007} Galperin, B., Sukoriansky, S., Anderson, P. S., 2007: On the critical Richardson number in stably stratified turbulence. \emph{Atm. Sci. Lett.}, \textbf{8} (3), 65-69.

\bibitem{Hanazaki} Hanazaki, H., and Hunt, J. C. R., 2004: Structure of unsteady stably stratified turbulence with mean shear. \emph{J. Fluid Mech.}, \textbf{507}, 1-42.

\bibitem{PT} Hauf, T., and H\"{o}ller, H.: 1987, Entropy and Potential
 Temperature, \textit{J. of Atm. Sci.}, \textbf{44}, 2887-2901.

\bibitem{Hunt88} Hunt, J. C. R., Stretch, D. D., and Britter, R. E., 1988: Length scales in stably stratified turbulent flows and their use in turbulence models. In: \emph{Proc. I.M.A. Conference on "Stably Stratified Flow and Dense Gas Dispersion"} (J. S. Puttock, Ed.), Clarendon Press, 285-322.

\bibitem{Keller00} Keller, K., and Van Atta, C. W., 2000: An experimental investigation of the vertical temperature structure of homogeneous stratified shear turbulence, \emph{J. Fluid Mech.}, \textbf{425}, 1-29.

\bibitem{K41} Kolmogorov, A. N., 1941: Energy dissipation in locally isotropic turbulence. Doklady AN SSSR, 32, No.1, 19-21.

\bibitem{Kurbatsky} Kurbatsky, A. F.: 2000, Lectures on Turbulence,
Novosibirsk State University Press, Novosibirsk.

\bibitem{LL} Landau, L.D., and Lifshitz, E.M.: 1987,
Course of Theoretical Physics:
Fluid Mechanics, Pergamon, New York, 552 pp.

\bibitem{Luyten02} Luyten, P. J., Carniel, S., and Umgiesser, G., 2002: Validation of turbulence closure parameterisations for stably stratified flows using the PROVESS turbulence measurements in the North Sea, \emph{J. Sea Research}, \textbf{47}, 239-267.

\bibitem{Lvov2006a} L'vov, V.S., Pomyalov, A., Procaccia, I., and Zilitinkevich, S.S., 2006a: Phenomenology of wall bounded Newtonian turbulence, \emph{Phys. Rev. E.}, \textbf{73}, 016303.

\bibitem{Lvov2006b} L'vov, V.S., Procaccia, I., and Rudenko O., 2006b: Analytic Model of the Universal Structure of Turbulent Boundary Layers, \emph{JETP Letters}, \textbf{84}, 67-73.

\bibitem{Lvov2008} L'vov, V.S., Procaccia, I., and Rudenko O., 2008: Universal Model of Finite Reynolds Number Turbulent Flow in Channels and Pipes, \emph{Phys. Rev. Lett.}, \textbf{100}, 054504.

\bibitem{MY74} Mellor, G. L., and Yamada, T., 1974: A hierarchy of turbulence closure models for planetary boundary layer, \emph{J. Atmos. Sci.}, \textbf{31}, 1791-1806.

\bibitem{Monin1954} Monin, A. S., and Obukhov, A. M., 1954: Main characteristics of the turbulent mixing in the atmospheric surface layer, \emph{Trudy Geophys. Inst. AN. SSSR}, \textbf{24}(151), 153-187.

\bibitem{1879Obe} Oberbeck, A.: 1879, \"{U}ber die W\"{a}rmeleitung
der Fl\"{u}ssigkeiten bei Ber\"{u}cksichtigung der Str\"{o}mung
infolge Temperaturdifferenzen, \emph{Ann. Phys. Chem.} (Leipzig)
\textbf{7}, 271-292.

\bibitem{Pope} Pope, S.B.: 2001, Turbulent Flows, Cambridge University Press, 771 pp.

\bibitem{Rahmann05} Rehmann, C. R., and Hwang, J. H., 2005: Small-scale structure of strongly stratified turbulence, \emph{J. Phys. Oceanogr.}, \textbf{32}, 154-164.

\bibitem{Richardson} Richardson, L. F., 1920: The supply of energy from and to atmospheric eddies. \emph{Pros. Roy. Soc. London}, \textbf{A 97}, 354-373.


\bibitem{Rotta51} Rotta, J. C., 1951: Statistische theorie nichthomogener turbulenz , \emph{Z. Physik}, \textbf{129}, 547-572.

\bibitem{Schuman95} Schumann, U., and Gerz, T., 1995: Turbulent mixing in stably stratified sheared flows. \emph{J. Applied Meteorol.}, \textbf{34}, 33-48.

\bibitem{Stretch01} Stretch, D. D., Rottman, J. W., Nomura, K. K., and Venayagamoorthy, S. K., 2001: Transient mixing events in stably stratified turbulence, In: \emph{14th Australasian Fluid Mechanics Conference}, Adelaide, Australia, 10-14 December 2001.

\bibitem{Umlauf} Umlauf, L., and Burchard, H., 2005: Second-order turbulence closure models for geophysical boundary layers. A review of recent work. \emph{Continental Shelf Research}, \textbf{25}, 725-827.

\bibitem{Wyngaard} Weng, W., and Taylor, P., 2003: On modelling the one-dimensional Atmospheric Boundary Layer, \emph{Boundary-layer Meteorology}, \textbf{107}, 371-400.

\bibitem{Wyngaard} Wyngaard, J.: 1992, Atmosferic turbulence, \emph{Ann. Rev. Fluid Mech.} \textbf{24}, 205-233.

\bibitem{Zeman} Zeman, O.: 1981, Progress in the modeling of planetary boundary layers,  \emph{Ann. Rev. Fluid Mech.} \textbf{13}, 253-272.

\bibitem{Zil-IGW} Zilitinkevich S.S., 2002: Third-order transport due to internal
waves and non-local turbulence in the stably stratified surface
layer, \emph{Quarterly Journal of the Royal Meteorological Society},
\textbf{128}, 913-925.

\bibitem{HB} Zilitinkevich, S.S., Elperin, T., Kleeorin, N., and Rogachevskii, I., 2007: Energy- and flux-budget (EFB) turbulence closure model for stably stratified flows. Part I: steady-state, homogeneous regimes, \emph{Boundary-layer Meteorology} \textbf{125}, 167-191.
\bibitem{ZE} Zilitinkevich, S.S., and Essau, I.:  Similarity theory and calculation of turbulent fluxes at the surface for the stably stratified atmospheric boundary layer, \emph{Boundary-Layer Meteorology} \textbf{125}, 193-205 (2007).


\end{thebibliography}
\end{document}